\documentclass[reprint,amsmath,amssymb,aps,pra,superscriptaddress,floatfix]{revtex4-1}
\usepackage{graphicx}
\graphicspath{{figs/}}
\usepackage{bm}
\usepackage{hyperref}
\usepackage{xcolor}
\usepackage{stmaryrd}
\usepackage{esint}
\usepackage{relsize}
\usepackage{bbold}
\def\vec#1{\bm{#1}}
\def\abs#1{\left\lvert#1\right\rvert}

\let\Li\relax\DeclareMathOperator{\Li}{Li}
\allowdisplaybreaks

\begin{document}

\title{Quantum critical thermal transport in the unitary Fermi gas}

\author{Bernhard Frank}
\affiliation{Max-Planck-Institut f\"ur Physik komplexer Systeme, 01187
  Dresden, Germany}
\author{Wilhelm Zwerger}
\affiliation{Technische Universit\"at M\"unchen, Physik Department,
  James-Franck-Strasse, 85748 Garching, Germany}
\author{Tilman Enss}
\affiliation{Institut f\"ur Theoretische Physik,
  Universit\"at Heidelberg, 69120 Heidelberg, Germany}
\date{\today}

\begin{abstract}
  Strongly correlated systems are often associated with an underlying
  quantum critical point which governs their behavior in the finite
  temperature phase diagram. Their thermodynamical and transport
  properties arise from critical fluctuations and follow universal
  scaling laws.  Here, we develop a microscopic theory of thermal
  transport in the quantum critical regime expressed in terms of a
  thermal sum rule and an effective scattering time.  We explicitly
  compute the characteristic scaling functions in a quantum critical
  model system, the unitary Fermi gas.  Moreover, we derive an exact
  thermal sum rule for heat and energy currents and evaluate it
  numerically using the nonperturbative Luttinger-Ward approach.  For
  the thermal scattering times we find a simple quantum critical
  scaling form.  Together, the sum rule and the scattering time
  determine the heat conductivity, thermal diffusivity, Prandtl number
  and sound diffusivity from high temperatures down into the quantum
  critical regime.  The results provide a quantitative description of
  recent sound attenuation measurements in ultracold Fermi gases.
\end{abstract}
\maketitle

\section{Introduction}
Thermal transport caused by temperature gradients is ubiquitous in
nature and typically occurs in a diffusive manner. A calculation of
the corresponding thermal conductivity $\kappa$ and the associated
thermal diffusion constant $D_T=\kappa/c_p$ is often based on a
kinetic theory description like the Boltzmann equation.  This works
well, e.g., in metals at low temperature and allows one to understand
the origin of universal laws like the Lorenz ratio
$L=\kappa/\sigma T\to L_0=\pi^2k_B^2/3e^2$ between the thermal and the
electrical conductivity $\sigma$ as predicted by Wiedemann and Franz.
In strongly correlated systems, sometimes called bad metals
\cite{emery1995}, the underlying Fermi liquid description does not
apply, however, and $L$ deviates substantially from its ideal value
$L_0$ \cite{bruin2013, keimer2015}.  Developing a microscopic theory
for thermal transport in non-Fermi liquids has been a major challenge
for many years. It has been approached using a number of different
techniques like the memory function formalism \cite{mahajan2013}.  In
a number of cases, a possible and phenomenologically often successful
strategy to describe transport in the absence of well-defined
quasiparticles is to assume the existence of an underlying quantum
critical point (QCP) \cite{sachdev2011}.  Transport in the
quantum critical regime (QCR) above the QCP may then be
analyzed in terms of critical fluctuations where decay and scattering
rates typically scale linearly with temperature according to a
Planckian law $\tau^{-1} \propto k_BT/\hbar$~\cite{sachdev2011,
  hartnoll2018, lucas2019}, a behavior which has been observed
recently in the thermal diffusivity of near optimally doped cuprates
above the superconducting transition \cite{zhang2019}. The aim of our
present work is to develop a microscopic theory for thermal transport
in a much simpler system with a quantum critical point, namely the
unitary Fermi gas (UFG)~\cite{zwerger2012, zwer14varenna}. This system
has a QCP at zero density which is both scale and conformally
invariant~\cite{nikolic2007, sachdev2012, nishida2007nonrel,
  nishida2012}.  In the quantum critical regime above this point, the
thermal wavelength $\lambda_T=h/\sqrt{2\pi mT}$ (we set $k_B=1$
throughout the paper) and the characteristic time $\hbar/T$ are the
only relevant length and time scales. Correlation functions involving
observables that do not develop anomalous dimensions associated with
details of the interaction at short distances thus obey simple scaling
laws~\cite{zwer14varenna}.  This applies for instance to the shear
viscosity $\eta$ and the related ratio $\eta/s$ with the entropy
density $s$, which turns out to be not far above the well-known
Kovtun-Son-Starinets bound \cite{kovtun2005, enss2011, joseph2015,
  bluhm2017}. Similarly, the spin diffusion constant $D_s$ exhibits
the quantum critical scaling behavior, and a minimum value
$D_s\simeq 1.3\, \hbar/m$ has been measured and determined
theoretically~\cite{sommer2011a, enss2012spin, trotzky2015,
  valtolina2017}.
\begin{figure}[t]
  \centering
  \includegraphics[width=\linewidth]{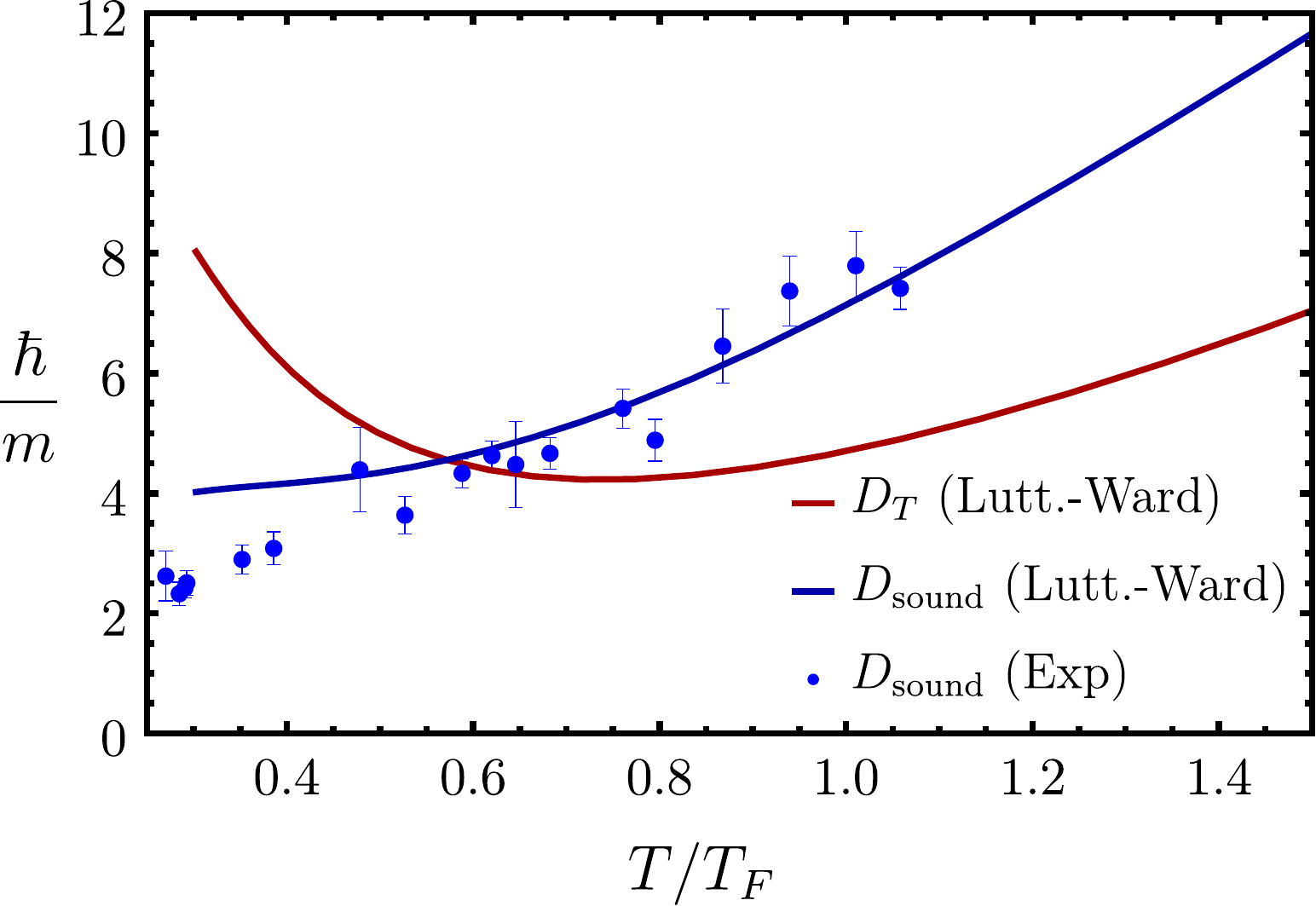}
  \caption{Thermal diffusivity $D_T$ (red) and sound diffusivity
    $D_\text{sound}$ (blue) vs temperature $T/T_F$ in the quantum
    critical regime of the unitary Fermi gas.  Theoretical results
    from Luttinger-Ward calculations are shown in comparison with
    sound diffusion measurements \cite{patel2019}.}
  \label{fig:diffusion}
\end{figure}

Very recently, experiments with dilute ultracold atomic gases have
realized homogeneous Fermi gases \cite{mukherjee2017, hueck2018} and
opened the possibility to access \emph{local} thermal transport via
the diffusive spreading of density and thermal wave packets
propagating in a sufficiently large box~\cite{hu2018, zhang2018,
  baird2019, patel2019, kuhn2020}.  These experiments are considerably
more sensitive than previous global transport measurements from trap
collective modes.  For instance, the measurements of the hydrodynamic
sound dispersion $\omega_q=c_sq - i D_\text{sound} q^2/2 + \dotsm$ in
a homogeneous unitary Fermi gas of $^6\text{Li}$ atoms at MIT
\cite{patel2019} provide both the speed of sound $c_s$ and the sound
diffusivity $D_\text{sound} = (4/3)D_\eta + (c_p/c_V-1)D_T$
\cite{forster1975}.  Knowledge of the kinematic viscosity
$D_\eta=\eta/(mn)$ \cite{enss2011, joseph2015} and the Landau-Placzek
ratio $\text{LP}=c_p/c_V-1$ (Fig.~\ref{fig:LP} below) gives then
access to the thermal diffusivity $D_T$ in the quantum degenerate gas,
see Fig.~\ref{fig:diffusion}.  Theoretical results for thermal
transport are so far available only at high temperature from the
virial expansion~\cite{braby2010}. It is the goal of this work to
compute thermal transport at low temperature and in particular in the
quantum critical regime.

In the following, we compute thermal transport in the quantum critical
region of the unitary Fermi gas based on a decomposition of the
thermal conductivity as a product
\begin{align}\label{eq:kappa_def}
  \kappa T = \chi^T_{qq} \, \tau_\kappa
\end{align}
of a nontrivial, thermodynamic sum rule $ \chi^T_{qq} $ for the heat
current response and a thermal scattering time $\tau_\kappa$ which can
formally be derived within a memory function approach, cf.\
Sec.~\ref{sec:QCresponse}.  We show that both factors of this
decomposition can be described by universal scaling forms which
smoothly connect the quantum critical to the high-temperature
regime, where a virial expansion for the thermodynamic properties and
a Boltzmann equation for the associated scattering time is applicable.
In Sec.~\ref{sec:sum-rule} we derive an exact expression for the
thermal sum rule $\chi^T_{qq}$ in terms of Green's functions with the
help of Ward identities for energy and particle number conservation.
Based on nonperturbative results for the Green's functions from a
fully self-consistent Luttinger-Ward computation~\cite{haussmann2007,
  frank2018universal, frankdiss} we evaluate $\chi^T_{qq}$
numerically.  We find a strong enhancement of spectral weight in the
quantum critical regime compared to the noninteracting gas which
reaches two orders of magnitude in the quantum critical regime just
above the superfluid transition.  In Sec.~\ref{sec:scatteringtimes} we
compute the thermal scattering time $\tau_\kappa$ of order $T/\hbar$
using a large-$N$ expansion. Quite unexpectedly, the time $\tau_\kappa$
extrapolates in a simple manner from the Boltzmann gas limit down into
the quantum critical regime.  In Sec.~\ref{sec:results}, we combine
the results for the sum rule with the scattering times in
Eq.~\eqref{eq:kappa_def} to predict the thermal transport coefficient
$\kappa$, the diffusivity $D_T$ shown in Fig.~\ref{fig:diffusion}, and
the Prandtl number $\text{Pr}$. In particular, we find good agreement
with the experimentally observed values in the quantum critical
regime.  We conclude with a discussion in Sec.~\ref{sec:discussion}.


\section{Quantum critical thermal transport}
\label{sec:QCresponse}

In this section we first define the quantum critical regime of the
unitary Fermi gas in part~\ref{sec:QCR}, and discuss the crossover to
classical critical behavior close to the finite-temperature superfluid
transition.  In part~\ref{sec:memory}, we discuss the formal structure
of how to compute thermal transport in linear response from the Kubo
formula and its evaluation within the memory function formalism.

\subsection{Quantum critical regime}
\label{sec:QCR}

\begin{figure}[t]
  \centering
  \includegraphics[width=\linewidth]{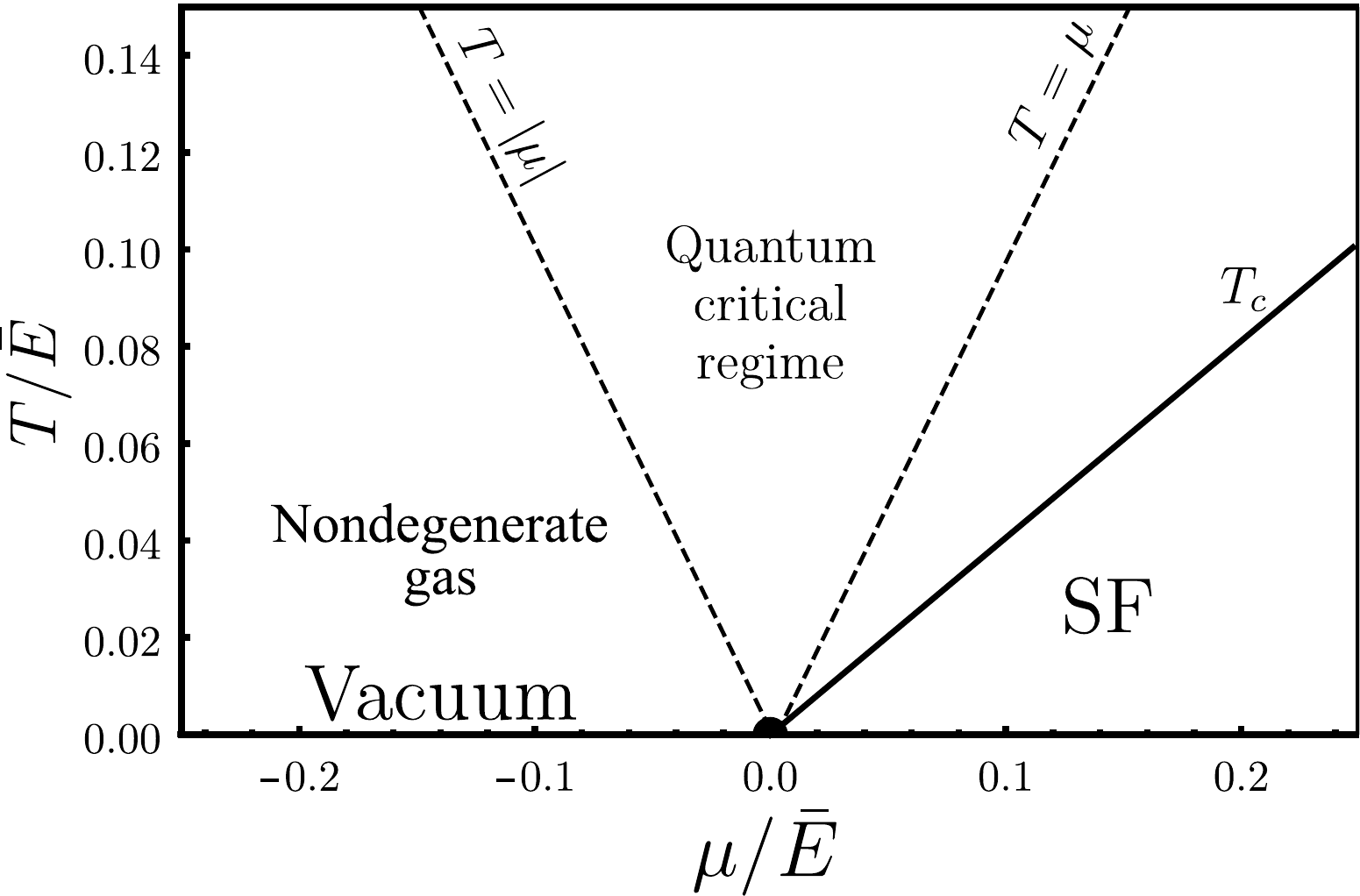}
  \caption{Phase diagram of the spin-balanced, unitary Fermi gas at
    finite temperatures~\cite{enss2012crit}.  The QCP at $T=0$,
    $\mu=0$ is the starting point for the phase boundary of the
    homogeneous superfluid at $T_c = 0.4 \mu$ (solid line). The dashed
    lines mark the crossover to the quantum critical regime above the
    QCP.}
  \label{fig:pd-UFG}
\end{figure}

Dilute ultracold Fermi gases interact via a short-range attractive
interaction between different spin
components~\cite{pethick_smith_2008}.  At low temperature, atoms
scatter predominantly in the $s$-wave channel with scattering
amplitude $f(k) = -1/(a^{-1} + ik)$, which is fully characterized by
the $s$-wave scattering length $a$.  Here we focus on the unitary
limit $1/a =0 $ that gives rise to a strongly interacting system as
the standard perturbative expansion in a small gas parameter
$n |a|^3 \ll 1$ breaks down.  The phase diagram shown in
Fig.~\ref{fig:pd-UFG} exhibits a quantum critical point at vanishing
chemical potential and temperature $\mu=T=0$, which separates the
vacuum state at $\mu<0$ from a homogeneous superfluid (SF) state at
$\mu >0$ \cite{nikolic2007, enss2012crit, zwer14varenna}.  Here, all
energies are expressed in terms of $\bar E$, which is of the order of
the van der Waals energy that sets the cutoff scale beyond which
details of the interaction potential start to matter.  The universal
description based on the model Hamiltonian \eqref{eq:singleC} below is
thus applicable only for $\mu, T \ll \bar E$.
In the absence of a finite effective Zeeman field
$h=(\mu_\shortuparrow - \mu_\shortdownarrow)/2$ which may lead to
nontrivial phases with a finite spin population
imbalance~\cite{frank2018universal, frankdiss} the phase diagram is
characterized by a single dimensionless parameter $\beta \mu$.  The
superfluid state remains stable for temperatures below the critical
curve $T_c \simeq 0.4 \mu$ or equivalently
$(\beta \mu)_c \simeq 2.5$~\cite{ku2012}.  Instead, for high
temperature or small fugacity $z=e^{\beta\mu}\ll1$ the system forms a
dilute, non-degenerate gas which can be described in terms of the
virial expansion.  Increasing the fugacity $z$ to values of order
unity one enters the QCR, as shown in Fig.~\ref{fig:pd-UFG}.  In this
regime, both thermodynamic and transport properties follow universal
scaling laws associated with the zero density fixed point at
$T=\mu=0$, with $\beta\mu$ as the single relevant scaling variable
\cite{nikolic2007}.

The quantum critical scaling is replaced by the one characteristic for
a classical $d=3$ XY model close to the superfluid phase transition at
$\mu_c(T)\simeq 2.5 T$.  This crossover occurs when the Gaussian
correlation length $\xi_G$ of the quantum model---defined by the
quadratic term $1/\xi_G^2=2m (\mu-\mu_c)/\hbar^2$ in the
Ginzburg-Landau free energy---becomes of the same order as the
characteristic length $\xi_1$.  The length $\xi_1\simeq 1/u_0$ is
associated with the coefficient $u_0$ of the quartic term
$(u_0/4!) \left( \phi_1^2(\mathbf{x})+ \phi_2^2(\mathbf{x}) \right)^2$
of the classical $\phi^4$ theory for a complex scalar field
$\phi(\mathbf x) = \phi_1(\mathbf x) + i \phi_2 (\mathbf x)$ that
depends only on space.  This term may in principle be derived from the
usual complex order parameter $\psi(\mathbf{x},\tau)$ for the
superfluid transition by integrating out all nonzero Matsubara
frequencies $\Omega_n\ne 0$. In explicit form, this has been worked
out for a generalization of the proper $N=2$ component model of a
weakly interacting Bose gas to a large number $N$, which yields
$u_0^{\rm BEC}(a)=96\pi^2 a/\lambda_T^2$ in the $N=\infty$
limit~\cite{baym2000}.  In the case of the unitary Fermi gas at
$1/a=0$, simple dimensional analysis requires that
$u_0 \simeq 1/\lambda_T$, however, the value of the numerical prefactor
is unknown.  Qualitatively, the crossover condition
$\xi_1\simeq \xi_G$ thus gives the simple relation
$\mu-\mu_c(T)\simeq k_BT$, which entails a Ginzburg parameter of order
unity and a very large Ginzburg region that extends up to about $2T_c$
as discussed by~\textcite{debelhoir2016}.

In the vicinity of the superfluid transition, the quantum critical
scaling of dynamical quantities is replaced by the classical dynamical
scaling.  In particular, the thermal conductivity of the UFG is
described within Model F~\cite{hohenberg1977} with dynamical critical
exponent $z=3/2$ for the superfluid transition in the universality
class of the $d=3$ XY model.  As shown by~\textcite{ferrell1967}, this
implies a divergent thermal conductivity
\begin{align}
  \label{eq:critical}
  \kappa \sim (T-T_c)^{-\nu/2} \sim (T-T_c)^{-1/3} \, ,
\end{align}
as $T \to T_c^+$, which diverges with an exponent close to $1/3$ since
$\nu \approx 2/3$.

\subsection{Linear response and memory function formalism for the
  thermal conductivity}
\label{sec:memory}

A formally exact expression which in principle allows to calculate
transport coefficients for an arbitrary form of the underlying
Hamiltonian is based on linear response theory and the resulting Kubo
formula.  In the special case of the thermal conductivity at external
momentum $\mathbf q=0$, it is convenient to consider the \emph{heat}
current density $\mathbf j^q$ \cite{kadanoff1963}, which is defined as
the energy current $\mathbf j^E$ at constant particle number,
\textit{i.e.}, with the enthalpy per particle $w/n$ times the number
current density $\mathbf j$ subtracted:
\begin{align}
  \mathbf j^q = \mathbf j^E - (w/n) \mathbf j = \mathbf j^E -
  (\mu+T\tilde s) \mathbf j\, .
\end{align}
Here, we have used the Gibbs-Duhem relation
$w = \varepsilon+p = \mu n+Ts$ and defined the entropy per particle
$\tilde s=s/n$. In standard hydrodynamic terms this corresponds to the
definition of the thermal conductivity via Fourier's law
$\mathbf{j}^E = - \kappa \nabla T$ in the absence of a particle
current.  Microscopically, the effect of a weak temperature gradient
is encoded in the equilibrium retarded heat current response function
from linear response theory,
\begin{align}\label{eq:lin-resp}
  \chi_{qq}(\omega) = \frac{i}{\hbar} \int_0^\infty dt\, e^{i \omega t}\,
  \int d^3 x\, \left\langle \left[ \hat{\bm{\jmath}}^q(\mathbf x,
  t), \hat{\bm{\jmath}}^q(\mathbf 0,0) \right] \right \rangle_{\text{eq}}\, ,
\end{align}
where we suppress the argument $\mathbf q = 0$ from now on.  The
retarded commutator in Eq.~\eqref{eq:lin-resp} defines a positive and
even spectral representation of the dynamic thermal conductivity
\begin{align}\label{eq:kappadef2}
  \kappa (\omega) T &= \frac{\operatorname{Im}\chi_{qq}(\omega)}{\omega}\,, &
  \kappa & = \lim_{\omega \to 0} \kappa (\omega) \, .
\end{align}
Since a fully microscopic evaluation of the frequency-dependent
response function in a strongly interacting many-body system is
impossible, it is necessary to reduce the problem by restricting
attention to the dc-response and a simplified procedure to evaluate
the characteristic time scale $\tau_\kappa$ defined in
Eq.~\eqref{eq:kappa_def}. Such a procedure is provided by the memory
function formalism. It has been used to determine the dynamical charge
conductivity of simple metals some time ago by~\textcite{goetze1972}
and it provides a systematic and unified description for the
derivation of hydrodynamic equations of motion in fluids, see, e.g.,
the textbook by~\textcite{forster1975}.  More recently, the method has
been applied successfully to calculate transport coefficients in
systems without well-defined quasiparticles~\cite{mahajan2013,
  hartnoll2014transport, hartnoll2018}.  It is based on a formal
expression for the Laplace transform of the relaxation function
\begin{align}
  \label{eq:memory}
  \phi_{AB}(z)
  & =\frac{\chi_{AB}(z)-\chi_{AB}^T}{iz} \\
  & =\chi_{AC}^T\left[\frac{1}{M(z)-iz\chi^T}\right]_{CD}\chi_{DB}^T \notag \, ,
\end{align} 
in terms of a matrix $\chi_{AB}^T$ of static thermodynamic
susceptibilities of slow variables $A,B$ \footnote{Note that for
  nonergodic variables, the thermodynamic susceptibility $\chi^T_{AB}$
  differs from the corresponding static one
  $\chi_{AB}(\omega=0)$. This results in relaxation functions that
  approach a nonzero limit as $t\to\infty$, a problem, which does not
  show up for thermal transport in the unitary Fermi gas.}  and an
associated frequency-dependent memory matrix $M_{AB}(z)$ (we assume
that the operators $A$ and $B$ have the same sign under time reversal,
otherwise an additional contribution appears in the
denominator). Provided that this matrix has a finite limit $M(0)$ at
vanishing frequency, this leads to an expansion
\begin{align}
  \label{eq:expansion}
  \chi_{AB}(z) & =\chi_{AB}^T+iz\chi_{AC}^T\tau_{CB}+\dotsm
\end{align} 
of the dynamical response function at low frequencies, which defines a
matrix of relaxation times
\begin{align}
  \tau_{CB} & =M^{-1}(0)_{CD}\chi_{DB}^T\,.
\end{align} 
Identifying $\kappa_{AB}T=\phi_{AB}(z=0)$ as the dc-transport
coefficient, this leads to $\kappa_{AB}T=\chi_{AC}^T\tau_{CB}$, which
is precisely of the form given in Eq.~\eqref{eq:kappa_def}.  In
principle, therefore, the memory function formalism determines
transport coefficients in quantum many-body systems in terms of the
matrix $\chi_{AB}^T$ of associated static thermodynamic
susceptibilities and the zero-frequency limit $M_{AB}(z=0)$ of the
memory matrix.  The formal expression for $M_{AB}(z)$ shows that it is
again a relaxation function but now for operators $\hat{Q}\dot{A}$ in
which the dynamics of the slow variables $A,B$ is projected out by
$\hat{Q}=1-\sum_{AB} (\chi^T)^{-1}_{AB}\vert A )\,(B\vert$.  In
practice, the memory matrix can hardly be determined exactly.  In
systems without long-lived quasiparticles, however, even approximate
results for the scattering times are often sufficient because the only
exactly or approximately conserved quantities are then particle
number, momentum and energy while all other variables relax on
microscopic time scales.

In fact, much of the nontrivial structure of transport coefficients
near quantum critical points is determined by the associated
thermodynamic susceptibilities, which is behind the success of the
memory function formalism in this context.  This turns out to be the
case also for the unitary Fermi gas studied here.  Indeed, as will be
shown in Sec.~\ref{sec:sum-rule}, the relevant susceptibility
$\chi^T_{qq}$ exhibits a rather strong dependence on the scaling
variable $\beta\mu$ (see Fig.~\ref{fig:sumrule} below), while the
scattering time in Fig.~\ref{fig:tau} evolves rather smoothly from the
high-temperature limit down into the quantum critical regime,
essentially extrapolating the result obtained from a Boltzmann
equation calculation.  A similar situation also applies to other
transport coefficients, such as the shear viscosity $\eta=p\tau_\eta$,
where the sum rule is given by the pressure $p$ \cite{enss2011}, or
particle transport, where an analog of the product form
\eqref{eq:kappa_def} for the thermal conductivity has also been found
to hold.

In the following, we will determine the thermodynamic susceptibility
$\chi_{qq}^T$ by a direct Green function approach, using an extension
of exact Ward identities first derived by \textcite{polyakov1969} in
the context of transport in the vicinity of a thermal critical point.
Since the heat current is an ergodic variable, the result must
coincide with the associated dc-susceptibility, which is given by the
standard Kramers-Kronig relation,
\begin{align}\label{eq:sumruledef}
  \chi^T_{qq} = \int \frac{d\omega}{\pi}
  \kappa(\omega, \mathbf q=0)_\text{reg}\,T \,,
\end{align}
as the integral of the frequency dependent heat conductivity
$\kappa(\omega, \mathbf q=0)$ times the temperature, including a
proper regularization of the divergences which arise as a result of
the assumption of a zero-range interaction in Eq.~\eqref{eq:singleC}
below.


\section{Thermal sum rule}
\label{sec:sum-rule}

In this section we first introduce the model for the interacting Fermi
gas in part~\ref{sec:model}, and we then express the linear response
theory for thermal transport in the field theoretical formulation
based on Green's functions (part \ref{sec:lin-resp}).  In particular,
we derive a new Ward identity for the interaction part of the heat
current, which gives rise to a novel exact expression for the thermal
sum rule \eqref{eq:sumrule}--\eqref{eq:sumrulepair} in terms of one-
and two-particle Green's functions.  Next we discuss in
part~\ref{sec:UV-reg} the necessary regularization of the
high-momentum asymptotics.  Finally, we numerically evaluate the sum
rule in the quantum critical regime using the nonperturbative
Luttinger-Ward approach in part~\ref{sec:num-res}.

\subsection{Model}
\label{sec:model}

The many-body physics of an ultracold Fermi gas with two trapped
hyperfine states (labeled $\shortuparrow,\shortdownarrow$) can be
described by the grand canonical Hamiltonian for spin-$1/2$ fermions
with contact interaction \cite{zwer14varenna},
\begin{multline}
\label{eq:singleC}
  \hat{H}=\int_{\mathbf{x}}\,\Big[
  \sum_{\sigma=\shortuparrow,\shortdownarrow}
  \hat{\psi}_{\sigma}^{\dagger} 
  \left(\mathbf{x}\right)
  \Big(-\frac{\hbar^2}{2m}\nabla^2-\mu_\sigma\Big)\hat{\psi}_{\sigma}(\mathbf{x})\;
  \\ 
  + \bar{g}(\Lambda)\,\hat{\psi}_{\shortuparrow}^{\dagger}(\mathbf{x})
  \hat{\psi}_{\shortdownarrow}^{\dagger}(\mathbf{x})
  \hat{\psi}_{\shortdownarrow}(\mathbf{x})
  \hat{\psi}_{\shortuparrow}(\mathbf{x})\Big]\, .
\end{multline} 
The total density is controlled by the chemical potentials, with
$\mu_\shortuparrow=\mu=\mu_\downarrow$ in the balanced case considered
here.  In order to reproduce a finite $s$-wave scattering amplitude
$f(k) = -1/(a^{-1} + ik)$, the strength of the effective contact
potential has to be chosen appropriately.  In three dimensions, this is
achieved in a standard manner by introducing a scale-dependent
coupling constant $\bar g(\Lambda)$, which is related to the physical
$s$-wave scattering length $a$ via
\begin{align}\label{eq:gbar}
  \bar g(\Lambda)^{-1} = m/(4\pi\hbar^2 a)-m\Lambda/(2\pi^2\hbar^2) \, .
\end{align}
Here, $\Lambda$ is a high-momentum cutoff whose physical origin is the
finite effective range $r_e$ of the actual interaction.  In the
experimentally relevant case of open-channel dominated Feshbach
resonances (e.g., for $^6\text{Li}$ at a magnetic field of 832
G~\cite{zuern2013}) the zero-range limit $\Lambda \to \infty$ is an
excellent approximation because $r_e\simeq \ell_\text{vdW}$ is of the
order of the van der Waals length $\ell_\text{vdW}$ and the associated
momentum scale $\Lambda\simeq \hbar/\ell_\text{vdW}$ is far beyond the
accessible range \cite{zwer14varenna}.  As a result, the momentum
distribution $n_\sigma(p\to\infty) = \mathcal C/p^4$ exhibits a
power-law behavior over a wide range of momenta with a strength
determined by the Tan contact density
$\mathcal C$~\cite{tan2008energetics}.  As will be discussed below,
similar power laws also appear in dynamical correlation functions.
The unitary limit $1/a =0$ can be reached by tuning the interaction
directly to the Feshbach resonance, which is controlled by an external
magnetic field.  As a result, there is no small interaction parameter
available and a nonperturbative treatment is mandatory to obtain
quantitative results.  The Luttinger-Ward approach results in
single-particle Green's functions $G_\sigma$ at finite temperature
with self-consistently resummed interaction effects and is in good
agreement with thermodynamic measurements in the strong-coupling
regime around the unitary limit \cite{ku2012, mukherjee2019}.

In addition to the fermionic Green's function
$G_\sigma(\mathbf x, \tau) = -\langle \mathcal T_\tau
\hat\psi_\sigma(\mathbf x, \tau) \hat\psi_\sigma^\dagger(\mathbf 0, 0)
\rangle$, the Luttinger-Ward theory also allows to determine the pair
propagator
\begin{multline}
  \Gamma(\mathbf x, \tau) = \bar{g}(\Lambda) \delta(\tau)
  \delta(\mathbf x) \\
  - \bar{g}(\Lambda)^2 \left \langle \mathcal T_\tau \left(\hat
        \psi_\shortuparrow \hat \psi_\shortdownarrow\right)(\mathbf x,
      \tau) \left(\hat \psi^\dagger_\shortdownarrow \hat
        \psi^\dagger_\shortuparrow\right)(\mathbf 0, 0) \right
  \rangle \, ,
\end{multline}
where $\mathcal T_\tau$ denotes time ordering in imaginary time
$\tau$.  At the superfluid transition temperature $T_c$, the pair
propagator $\Gamma(\mathbf Q=0,\Omega_n=0)$ diverges according to the
Thouless criterion.  Furthermore, the Tan contact is obtained from the
short-distance limit~\cite{haussmann2009}
\begin{align}
  \label{eq:contact}
  \frac{\hbar^4 \mathcal C}{m^2}- \Delta^2
  = - \Gamma(\mathbf x=0,\tau \to 0^-)\,,
\end{align}
where the anomalous contribution from the superfluid order parameter
$\Delta $ vanishes in the normal phase considered here. In the
following, both $G_\sigma$ and $\Gamma$ form the basis for the
evaluation of the thermal sum rule.

\subsection{Linear response}
\label{sec:lin-resp}

In order to determine the thermal conductivity of the UFG we first
evaluate the thermal sum rule $\chi_{qq}^T$.  In contrast to other
transport coefficients such as the viscosity \cite{enss2011} or the
spin diffusivity \cite{enss2012spin}, $\chi_{qq}^T$ cannot be directly
attributed to standard thermodynamic quantities but requires an
additional thermal operator~\cite{shastry2006}.

In general, the heat current response $\chi_{qq}(\omega)$ is obtained
within linear response by adding the perturbation
$\delta \hat H (t) = \int_\mathbf{x} \hat{\bm\jmath}^q(\mathbf{x},t)
\cdot \mathbf h(\mathbf x, t)$ to the Hamiltonian.  Rather than
working in real time, the problem is more conveniently treated in
imaginary time $\tau \in [0,\hbar\beta)$.  Furthermore, we consider
only homogeneous source terms
$\mathbf h(\mathbf x, \tau) = \mathbf h(\tau)$ since we are interested
in the $\mathbf q =0$ response.  We express the grand canonical
partition function in the presence of the external field $\mathbf h$
as a coherent state path integral~\cite{altland2006condensed} with fermionic action $S_F$,
\begin{subequations}
\begin{align}
  &\mathcal{Z}[\mathbf{h}]  = \int
    \prod_\sigma\mathcal{D}\bar{\psi}_\sigma  \mathcal{D}\psi_\sigma
    e^{-\frac{1}{\hbar} S_F[\bar{\psi}_\sigma ,\psi_\sigma, \mathbf{h}
    ]}\,, \\
&  S_F[\bar{\psi}_\sigma ,\psi_\sigma, \mathbf{h} ] =\int_0^{\hbar
    \beta}d\tau \sum_\sigma\int_\mathbf x \left(  \bar{\psi}_\sigma
    \partial_\tau \psi_\sigma \right)+ \notag \\
   &\qquad+\int_0^{\hbar \beta} d\tau
   \bigg( H[\bar\psi_\sigma,\psi_\sigma]+\bm j^q(\mathbf{q}=0, \tau) \cdot
    \mathbf h(\tau)\bigg)\, . \label{eq:action1}
\end{align}
\end{subequations}
Then $\log \mathcal Z$ is a generating functional for connected heat
current correlations
\begin{subequations}
\begin{align}\label{eq:mean-current}
  \left \langle \hat{\bm \jmath}^q(\mathbf q =0, \tau) \right \rangle
  & = -\left. \frac{\delta \log \mathcal{Z}[\mathbf h]}{\delta \mathbf h(\tau)}
    \right|_{\mathbf h= 0} \,,\\ 
  \chi_{qq}(\tau)
  & = \left. \frac{\delta^2 \log
                    \mathcal{Z}[\mathbf h] }{\delta \mathbf
                    h(\tau) \delta \mathbf h(0)} \right|_{\mathbf
                    h = 0} \,. \label{eq:response-derive}
\end{align}
\end{subequations}
From the latter function the retarded response in real frequency is
obtained by Fourier transforming $\tau$ to the bosonic Matsubara
frequency $\omega_n$ and subsequent analytic continuation
$i\omega_n \to \omega + i 0^+$.

\begin{figure}[t]
  \centering
  \includegraphics[width=.7\linewidth]{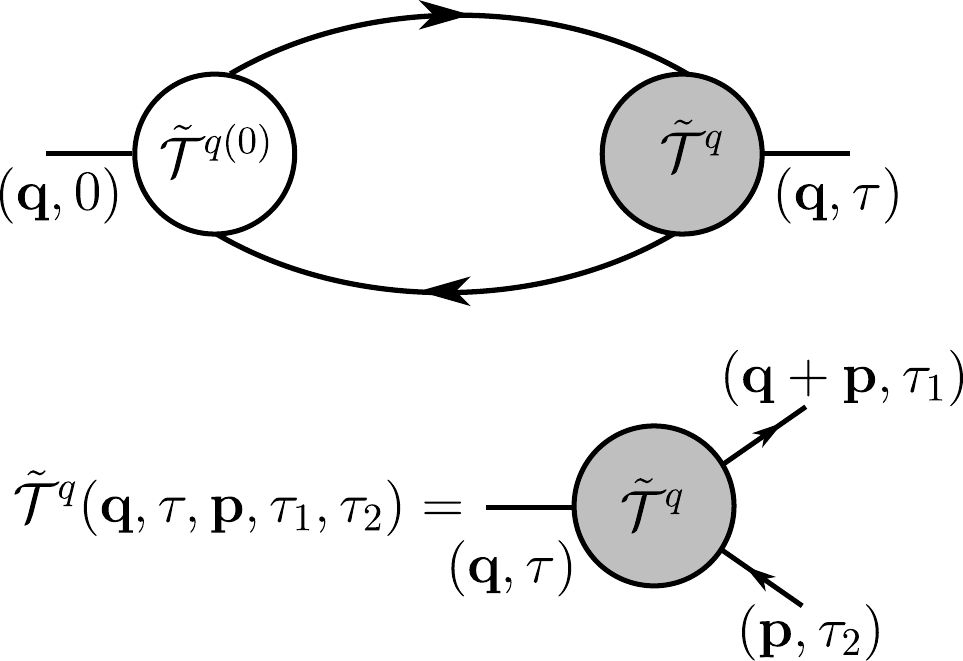}
  \caption{Diagrammatic representation of the current correlation
    function (Kubo formula) and the dressed, amputated current vertex
    $\Tilde{\mathcal T}^q$.  The total response is given by the sum of
    the fermionic and pair contributions.}
  \label{fig:Feynman}
\end{figure}

For the Hamiltonian~\eqref{eq:singleC} the particle and energy current
operators read \cite{forster1975, fujii2018}
\begin{subequations}\label{eq:currentsreal}
\begin{align}
  \hat{\bm{\jmath}}(\mathbf x)
  &= - \frac{i\hbar}{2m} \sum_\sigma (\psi^\dagger_\sigma\nabla \psi_\sigma(\mathbf x)
    -(\nabla\psi^\dagger_\sigma) \psi_\sigma(\mathbf x) )\,,  \\
  \hat{\bm{\jmath}}^E(\mathbf x)
  & = \hbar^3\sum_\sigma \frac{\nabla \psi_\sigma^\dagger \Delta \psi_\sigma - \Delta
    \psi_\sigma^\dagger \nabla \psi_\sigma}{4im^2} \\
  & \qquad \qquad + \frac{\hbar\bar g(\Lambda)}{im} \sum_{\sigma\neq\tau}
    \psi_\sigma^\dagger [\psi_\tau^\dagger
    \overset{\leftrightarrow}{\nabla} \psi_\tau]
    \psi_\sigma \notag \,.
\end{align}
\end{subequations}
The bare energy current operator $\hat{\bm \jmath}^E$ has a kinetic
and an interaction contribution. Considering the corresponding
operators in momentum space,
\begin{subequations}\label{eq:currents}
\begin{align}
  \hat{\bm \jmath}(\mathbf q =0 )
  &  = \sum_{\mathbf p \sigma}  \frac{\mathbf p}{m} c^\dagger_{\mathbf
    p,\sigma} c_{\mathbf p,\sigma}  \,,\\
  \hat {\bm \jmath}^E( \mathbf q=0)
  & = \sum_{\mathbf p\sigma} \frac{\mathbf p}{m}  \varepsilon_p
    c_{\mathbf p\sigma}^\dagger c_{\mathbf p\sigma} \\
  & \hspace*{-10mm} + \bar g(\Lambda) \sum_{\mathbf Q \mathbf p \mathbf p'} \frac{\mathbf Q}m
    c_{\mathbf Q/2+ \mathbf p\uparrow}^\dagger c_{\mathbf Q/2-\mathbf p\downarrow}^\dagger
    c_{\mathbf Q/2 + \mathbf p' \downarrow} c_{\mathbf Q/2 - \mathbf p'\uparrow} \, ,\notag 
\end{align}
\end{subequations}
shows that the prefactor of the interaction part is only sensitive to
the center-of-mass momentum $\mathbf Q$ of the pair of fermions
participating in the interaction. Therefore, this term is most easily
discussed in two-channel variables with a bosonic pair field
$\Delta(\mathbf x)=\bar g\psi_\downarrow(\mathbf
x)\psi_\uparrow(\mathbf x)$. The latter can be easily introduced by
decoupling the action~\eqref{eq:action1} in the pairing channel by a
Hubbard-Stratonovich transformation. We notice that the presence of
$\mathbf h(\tau)$ leads to the shift
$\varepsilon_p \to \varepsilon_p + (\varepsilon_p -\mu_\sigma -
T\tilde{s})\mathbf{p}/m \cdot \mathbf h(\tau)$ of the bare fermionic
dispersion relation and to the rescaling
$\bar g(\Lambda) \to \bar{g}(\Lambda) (1+ \mathbf Q/m \cdot \mathbf h
(\tau))$ in $S_F$.  With these substitutions we obtain the path
integral $\mathcal{Z}[\mathbf h]$ within the two-channel formulation
in momentum space,
\begin{widetext}
\begin{subequations}
\begin{align}
  \mathcal{Z}[\mathbf{h}]
  & = \int \prod_{\sigma}\mathcal{D}\bar{c}_{\sigma}  \mathcal{D}c_{
    \sigma} \mathcal{D} \bar\Delta \mathcal D \Delta \,
    e^{-\frac{1}{\hbar} S_{BF}[\bar{c}_{\sigma}
    ,c_{\sigma},\bar\Delta ,\Delta , \mathbf{h} ]} \,,\\
  \label{eq:action2}
  S_{BF}[\bar{c}_\sigma ,c_\sigma,\bar\Delta ,\Delta , \mathbf{h} ]
  &= \int_0^{\hbar \beta} d\tau \Bigl[\sum_{\mathbf p, \sigma}
    \bar{c}_{\mathbf p ,\sigma}(\tau)\bigl(\partial_\tau
    +\varepsilon_p  -\mu_\sigma + (\varepsilon_p-\mu_\sigma -
    T\tilde{s}) \frac{\mathbf p}{m}\cdot \mathbf h
    (\tau)\bigr) c_{\mathbf{p} \sigma}(\tau) \\
  & \qquad - \frac{1}{\bar{g}(\Lambda)} \sum_{\mathbf Q}
    \bigl(1 + \frac{\mathbf Q}{m} \cdot \mathbf h(\tau) \bigr)^{-1}
    \bar\Delta_{\mathbf Q}(\tau) \Delta_{\mathbf Q}(\tau) -
    \sum_{\mathbf p_1, \mathbf p_2} \left( \bar\Delta_{\mathbf p_1 +
    \mathbf p_2} (\tau) c_{\mathbf p_1\downarrow}(\tau)
    c_{\mathbf p_2 \uparrow}(\tau) + \text{h.c.} \right)\Bigr]\, .\notag 
\end{align}
\end{subequations}

From Eq.~\eqref{eq:mean-current}, we thus find the expectation value
of the heat current
\begin{align}\label{eq:jcurrent}
  \left \langle \hat{\bm\jmath}^q (\mathbf q=0, \tau) \right \rangle
  = \sum_{\mathbf p\sigma} \tilde{\bm{\mathcal{T}}}^{q(0)}_{\sigma} (\mathbf p)
  G_\sigma (\mathbf p , \tau - \tau^+) - \sum_\mathbf{Q}
  \tilde{\bm{\mathcal{T}}}_\text{pair}^{q(0)}(\mathbf Q)
  \Gamma(\mathbf Q, \tau - \tau^+) \, ,
\end{align}
\end{widetext}
where we have defined the bare fermionic and bosonic heat current vertices
\begin{subequations}\label{eq:barevertex}
\begin{align}
  \tilde{\bm{\mathcal{T}}}^{q(0)}_\sigma(\mathbf p)
  & = \left(\varepsilon_p - \mu_\sigma -T\tilde{s}\right)
    \frac{\mathbf p}{m} \,,\\ 
  \tilde{\bm{\mathcal{T}}}^{q(0)}_\text{pair}(\mathbf Q)
  & = \frac{1}{\bar{g}(\Lambda)}\frac{\mathbf Q}{m} \, ,
\end{align}
\end{subequations}
and furthermore recovered the single-particle Green's function
$G_\sigma$ as well as the pair propagator $\Gamma$ defined above.  A
vanishing perturbation $\mathbf h(\tau) = 0$ implies
$\left \langle \hat{\bm \jmath}^q (\mathbf q=0, \tau) \right
\rangle=0$ by rotation invariance.  Next, we obtain the susceptibility
by taking the second order functional derivative according to
Eq.~\eqref{eq:response-derive},
\begin{align}
\begin{split}
  \chi_{qq}(\tau) =
  & -\sum_{\mathbf p \sigma} \tilde{\bm{ \mathcal{T}}}^{q(0)}_{\sigma} (\mathbf
  p) \bm{ \mathcal{T}}_{\sigma}^q (\mathbf 0, \tau , \mathbf p, 0, 0^+)
  \\
  & + \sum_\mathbf{Q} \tilde{\bm{\mathcal{T}}}_\text{pair}^{q(0)}(\mathbf Q)
  \bm{\mathcal{T}}_\text{pair}^q (\mathbf 0, \tau , \mathbf Q, 0, 0^+)
  \\
  & -\frac{2 \delta(\tau)}{3m\bar{g}(\Lambda)} \sum_{\mathbf{Q}}
  \frac{\mathbf Q^2}{m} \Gamma(\mathbf Q ,0,0^+)
  \mathlarger{\mathlarger{\mathbb{1}}}_{3 \times 3}\, .
\end{split}
\end{align}
Here, the dressed current vertices for spin component $\sigma$ and the
pairs are defined as the time-ordered expectation values
\begin{subequations}\label{eq:vertex}
\begin{align}
  \bm{\mathcal{T}}^q_\sigma(\mathbf q, \tau, \mathbf p, \tau_1, \tau_2)
  &= \langle \mathcal T_\tau \hat{\bm{\jmath}}^q (\mathbf q, \tau)
    c_{\mathbf{p}+\mathbf q \sigma}(\tau_1 ) c_{\mathbf p
    \sigma}^\dagger(\tau_2) \rangle \notag \\ 
  & = \left. \frac{\delta G_\sigma(\mathbf p, \tau_1, \tau_2)}{\delta
    \mathbf h(\mathbf q ,\tau)}\right|_{\mathbf h =0}  \\
  \noalign{\vskip 10pt}
  \bm{\mathcal{T}}^q_\text{pair}(\mathbf q, \tau, \mathbf Q, \tau_1,
  \tau_2)
  &= \langle \mathcal T_\tau \hat{\bm{\jmath}}^q (\mathbf q, \tau)
    \Delta_{\mathbf{Q}+\mathbf q }(\tau_1 ) \Delta_{\mathbf
    Q}^\dagger(\tau_2) \rangle  \notag \\
  & = \left. \frac{\delta \Gamma(\mathbf Q, \tau_1, \tau_2)}{\delta
    \mathbf h(\mathbf q ,\tau)}\right|_{\mathbf h =0}  \, ,
\end{align}
\end{subequations}
while the last contribution to $\chi_{qq}$ arises from the second
derivative of the $\bar\Delta \Delta$ prefactor in the
action~\eqref{eq:action2}.  The thermal sum rule follows by Fourier
transformation to the external bosonic Matsubara frequencies
$\omega_n=0$, which yields
\begin{align}\label{eq:chiWn}
\begin{split}
  \chi_{qq}(\omega_n=0)
  =& -\sum_{\mathbf p \sigma, \epsilon_m} \tilde{\bm{\mathcal{T}}}^{q(0)}_{\sigma} (\mathbf p)
  \bm{ \mathcal{T}}^q_{\sigma} (\mathbf 0, \omega_n=0 , \mathbf p, \epsilon_m) \\
  &   + \sum_{\mathbf{Q},\Omega_m} \tilde{\bm{\mathcal{T}}}_\text{pair}^{q(0)}(\mathbf Q)
  \bm{\mathcal{T}}^q_\text{pair} (\mathbf 0, \omega_n=0, \mathbf Q, \Omega_m) \\
  & -\frac{2}{3m\bar{g}(\Lambda)} \sum_{\mathbf{Q},\Omega_m}
  \frac{\mathbf Q^2}{m} \Gamma(\mathbf Q ,\Omega_m)
  \mathlarger{\mathlarger{\mathbb{1}}}_{3 \times 3}\, .
\end{split}
\end{align}
This form of the sum rule in terms of current vertex functions is
analogous to the sum rules for momentum \cite{taylor2010, enss2011,
  enss2019bulk} and spin currents \cite{enss2013sum}. Introducing
$\tilde{\bm{\mathcal T}}^q_{\sigma, \text{pair}}$ as the amputated
counterparts of $\bm{\mathcal T}^q_{\sigma, \text{pair}}$ allows one
to represent the Kubo formula for $\chi_{qq}$ and thus also the sum
rule in a diagrammatic manner, as depicted in Fig.~\ref{fig:Feynman},
except for the last line.

Quite crucially, the exact heat current vertex $\tilde{\mathcal{T}}^q$
satisfies a Ward identity~\cite{polyakov1969}.  Extending the latter
from the fermionic to the bosonic sector, it reads in momentum space
(cf.\ Appendix~\ref{sec:ward})
\begin{subequations}\label{eq:ward}
\begin{align}
  &\tilde{ \mathcal T}^q_{\sigma}(\mathbf p,\varepsilon)
  = (T\tilde s-\varepsilon)\frac{\partial
    G_\sigma^{-1}(\mathbf p,\varepsilon)} {\partial \mathbf p}
    - \frac{\mathbf p}m G_\sigma^{-1}(\mathbf p,\varepsilon), \\
  & \tilde {\mathcal T}^q_{\text{pair}}(\mathbf Q,\Omega)
    = -\Omega \frac{\partial\Gamma^{-1}(\mathbf
    Q,\Omega)}{\partial \mathbf Q}
    + \frac{\mathbf{Q}}{m} \Gamma^{-1}(\mathbf Q,\Omega) \, ,
\end{align}
\end{subequations}
at vanishing external arguments $\omega=0$, $\mathbf q \to \mathbf 0$
relevant for the sum rule.  The first line contains the fermionic part
expressed via the single-particle Green's function $G_\sigma$, while
the second line denotes the bosonic contribution in terms of the pair
propagator $\Gamma$.  The \emph{bare} vertices
$\tilde{\bm{\mathcal{T}}}_\sigma^{q(0)}$ and
$\tilde{\bm{\mathcal{T}}}_\text{pair}^{q(0)}$ in
Eq.~\eqref{eq:barevertex} are obtained simply by using the
noninteracting Green function
$G_{0,\sigma}^{-1}(p,\varepsilon) = \varepsilon-\varepsilon_p+\mu$ and
the inverse bare coupling
$\Gamma_0^{-1}(Q,\Omega) = \bar g(\Lambda)^{-1}$ inside the Ward
identity (see Appendix~\ref{sec:ward}).
  
In order to obtain the sum rule as the static limit of the current
response function given by the Kubo formula (Fig.~\ref{fig:Feynman})
we insert the Ward identities into Eq.~\eqref{eq:chiWn}.  As a result,
we obtain the \emph{exact} thermal sum rule expressed in terms
  of the Green's and vertex functions,
\begin{align}
  \label{eq:sumrule}
  \bar\chi_{qq}^T(T,\mu)
  =\bar\chi_{qq}^{T,\text F}(T, \mu) + \bar\chi_{qq}^{T,\text{pair}}(T, \mu),
\end{align}
with two contributions: a fermionic part 
\begin{multline}
  \label{eq:sumruleF}
  \bar\chi_{qq}^{T,\text F}
  = -\frac1{\beta V} \sum_{p\sigma\epsilon_n} \frac{p^2}{3m^2}
  (\varepsilon_p-\mu-T\tilde s) \\
  \times [(i\epsilon_n-T\tilde s)\frac mp
  \frac{\partial G_\sigma(p,i\epsilon_n)}{\partial p} -
  G_\sigma(p,i\epsilon_n)]
\end{multline}
and a new interaction part arising from the bosonic pairs of the form
\begin{multline}
  \label{eq:sumrulepair}
  \bar\chi_{qq}^{T,\text{pair}}
  = \Bigl(\frac{m}{4\pi\hbar^2a}-\frac{m\Lambda}{2\pi^2\hbar^2}\Bigr)\\
  \times \frac{1}{\beta V} \sum_{Q\Omega_n}
  \frac1m \Bigl(\frac{Q^2}{3m}-i\Omega_n\Bigr)
  \Gamma(Q,i\Omega_n)  \, .
\end{multline}
Both terms can be evaluated by inserting the previously computed
Luttinger-Ward results for $G_\sigma(p, i \varepsilon_n)$ and
$\Gamma(Q, i \Omega_n)$ \cite{haussmann2007, frank2018universal,
  frankdiss} as functions of momentum $p$ ($Q$) and Matsubara
frequency $i\varepsilon_n$ ($i\Omega_n$).

Note that the full fermionic and bosonic energy current vertices as
defined by the Ward identity \eqref{eq:ward} provide an exact solution
of the Luttinger-Ward transport equations formulated in terms of
fermionic and bosonic transport vertices \cite{enss2011,
  enss2019bulk}.  This proves that the Luttinger-Ward approach
implements \emph{exact} energy conservation, even when fermionic and
bosonic Green functions are obtained within the self-consistent
T-matrix approximation.  This was indeed the goal of constructing a
conserving approximation, which in our case furthermore satisfies the
exact Tan relations \cite{enss2012crit}.
  
However, as indicated by the bar, these terms still depend explicitly
on the momentum cutoff $\Lambda$, which is manifest for the second
term.  Moreover, a finite value of $\Lambda$ is necessary to render
the momentum integrals in the fermionic part finite.  Therefore, we
first have to discuss how to extract the universal results for the sum
rule before presenting the numerical results.

\subsection{Short-distance asymptotics}
\label{sec:UV-reg}

Due to the contact interaction, several terms in the sum
rule~(\ref{eq:sumruleF},\ref{eq:sumrulepair}) diverge in the
zero-range limit $\Lambda\to\infty$.  This is apparent for the pair
contribution, which in the unitary limit $1/a=0$ is directly
proportional to $\Lambda$.  Indeed, one quite generally expects a
cutoff dependence of the static sum rules for these quantum critical
systems.  This can be attributed to high-frequency tails of the
dynamic transport coefficients such as $\kappa(\omega)$ defined in
Eq.~\eqref{eq:kappadef2} above \cite{hartnoll2018}.  For instance, in
case of the shear viscosity the full sum rule reads \cite{taylor2010,
  enss2011}
\begin{align}
  \label{eq:sumrulevisc}
  \bigl \langle \hat \Pi_{xy} \hat \Pi_{xy} \bigr \rangle_{\omega = 0}
  = p + \frac{4\hbar^2\mathcal C \Lambda}{15\pi^2m}\,,
\end{align}
where the second term arises from the high-frequency tail
$\eta(\omega \to \infty) = \hbar^{3/2}\mathcal C
/15\pi\sqrt{m\omega}$.  The static transport coefficient $p$ is given
instead by the regularized form of the sum rule with the
$\Lambda$-dependent terms subtracted.  As a result for the shear
viscosity of the unitary gas, one has $\eta = p\tau_\eta$ in analogy
to Eq.~\eqref{eq:kappa_def} in the thermal case.

These divergences arise from the asymptotic large-momentum behavior of
the fermionic momentum distribution,
\begin{align}
  \label{eq:Tan-momentum}
  n_\sigma(p\to\infty)
  \sim \mathcal C/p^4 + \mathcal D_1/p^6 + \mathcal D_2/p^7 + \dotsm \, , 
\end{align}
where we identify a new contribution $\mathcal D_2/p^7$ that is
clearly seen in our numerical data.  As discussed in
Appendix~\ref{sec:UV-LW}, its origin may be traced back to a
next-to-leading order non-analytical contribution
$\Gamma(\mathbf{x}=\mathbf 0, \tau \to \beta^-) \sim
(\beta-\tau)^{3/2}$ in the pair propagator at short times.  The
appearance of the two leading contributions in the asymptotic
power-law decay of the momentum distribution arises from the
nonanalytic contributions proportional to $\vert\mathbf{x}\vert$ and
$\vert\mathbf{x}\vert^3$, respectively, in the short-distance operator
product expansion
\begin{multline}
  \label{eq:OPE}
  \hat\psi_{\sigma}^{\dagger} (\mathbf{R}+\frac{\mathbf{x}}{2})
  \hat\psi_{\sigma} (\mathbf{R}-\frac{\mathbf{x}}{2})
  = \hat{n}_{\sigma}(\mathbf{R})
  +i\hbar^{-1}\mathbf{x}\cdot\hat{\mathbf{p}}_{\sigma}(\mathbf{R}) \\
  -\frac{\vert\mathbf{x}\vert}{8\pi}\,\hbar^{-4}m^2\bar{g}^2(\Lambda)
  \,\hat\psi_{\uparrow}^{\dagger}\hat\psi_{\downarrow}^{\dagger}
  \hat\psi_{\downarrow}\hat\psi_{\uparrow}(\mathbf{R}) \\
  +\frac{\vert\mathbf{x}\vert^3}{96\pi}\,\hbar^{-4}m^2\bar{g}^2(\Lambda)
  \nabla^2_{\mathbf{R}}
  \,\hat\psi_{\uparrow}^{\dagger}\hat\psi_{\downarrow}^{\dagger}
  \hat\psi_{\downarrow}\hat\psi_{\uparrow}(\mathbf{R})+\dotsm
\end{multline}
of the one-particle density matrix~\cite{braaten2008}.  The
coefficients $\mathcal C$ and $ \mathcal D_1$ in
Eq.~\eqref{eq:Tan-momentum} are defined through the expectation values
of the contact operator
$\Hat{\mathcal C}(\mathbf{R})=\hbar^{-4}m^2\bar{g}^2(\Lambda)\,
\hat\psi_{\uparrow}^{\dagger} \hat\psi_{\downarrow}^{\dagger}
\hat\psi_{\downarrow} \hat\psi_{\uparrow}(\mathbf{R})$ and its second
derivative $\nabla^2_{\mathbf{R}}\,\Hat{\mathcal C}(\mathbf{R})$; note
that in $d=3$ the Fourier transform of $\vert\mathbf x\vert$ is
$-8\pi/p^4$ while $\vert\mathbf x\vert^3$ gives $96\pi/p^6$.  The
presence of a subleading term $\mathcal D_1/p^6$ in the momentum
distribution of two-component Fermi gases has been discussed in detail
by \textcite{werner2012}.  In general, the coefficient $\mathcal D_1$
also contains a contribution which involves the derivative of the
energy with respect to the effective range of the interaction.  In our
model, no such contribution appears and the full expression for
$\mathcal D_1$ is given in terms of the first-order time and
second-order spatial derivative of the pair propagator, see
Eqs.~\eqref{eq:coeffGamma} and \eqref{eq:momdistasymp} in
Appendix~\ref{sec:UV-LW}.

Within the self-consistent T-matrix approximation to the
Luttinger-Ward functional the powers of momentum are correctly
reproduced, whereas the contact coefficients
$\mathcal C, \mathcal D_1$ and $\mathcal D_2$ that characterize the
short-distance correlations as functions of $T, \mu$ and $a^{-1}$ in
the many-body medium are obtained approximately.  The asymptotic
behavior of the numerical data of the fermionic momentum distribution
is consistent both with the OPE and Ref.~\cite{werner2012} up to
$p^{-6}$, but to our knowledge the $p^{-7}$ contribution has not been
discussed before.  The latter arises from an anomalous contribution to
the pair propagator
$\Gamma(\mathbf{x}=\mathbf 0, \tau \to \beta^-) \sim
(\beta-\tau)^{3/2}$, see App.~\ref{sec:UV-LW}.
  
In the fermionic part \eqref{eq:sumruleF}, the leading divergence
$\mathcal O(\Lambda^3)$, which could arise from the $\mathcal C/p^4$
tail of the momentum distribution, cancels between the first and last
term in the square brackets, hence there is no $\Lambda^3$ divergence.
According to Eq.~\eqref{eq:Tan-momentum}, this leaves terms of order
$\mathcal O(\Lambda)$ and $\mathcal O (\log(\Lambda/\bar k))$, where
$\bar k$ denotes the momentum scale beyond which the algebraic power
laws of the terms in $\chi_{qq}^T$ dominate; in practice, one has
$\bar k \gtrsim 10 /\lambda_T$.  The coefficients of these subleading
divergences depend on $\mathcal C$, $\mathcal D_1$, and $\mathcal D_2$
(for the log term).  Similarly, for the pair momentum distribution we
find the asymptotic expansion
$n_\text{pair}(Q\to\infty) = 64\pi^2n\mathcal C/3Q^6 + \dotsm$, see
Eq.~\eqref{eq:Gamma-asy} in App.~\ref{sec:UV-LW}.  This implies that
the momentum sum in the interaction term \eqref{eq:sumrulepair} is
finite, while the inverse bare coupling in the prefactor diverges as
$\mathcal O(\Lambda)$.  In the numerical evaluation, we subtract all
divergent terms to obtain the regularized sum rule
\eqref{eq:sumruledef}, as has been done for the shear viscosity
\cite{enss2011}.  At unitary, in particular, the interaction term does
not contribute to the regularized sum rule as its contribution scales
like $1/a$. Away from unitarity $1/a\neq0$, in turn, the bosonic part
gives rise to a new contact correlation contribution to the thermal
conductivity similar to what has been found in the bulk viscosity
\cite{enss2019bulk, nishida2019, hofmann2020}.  Regarding the dynamic
thermal conductivity, the $\mathcal O(\Lambda)$ contribution implies a
tail $\kappa(\omega \to \infty) \sim \omega^{-1/2}$, while
$\mathcal O(\ln\Lambda)$ causes a subleading contribution to the
high-frequency behavior proportional to $\omega^{-1}$, in analogy to
the discussion below Eq.~\eqref{eq:sumrulevisc}.

\subsection{Numerical Results}
\label{sec:num-res}

\begin{figure}[t]
  \centering
  \includegraphics[width=\linewidth]{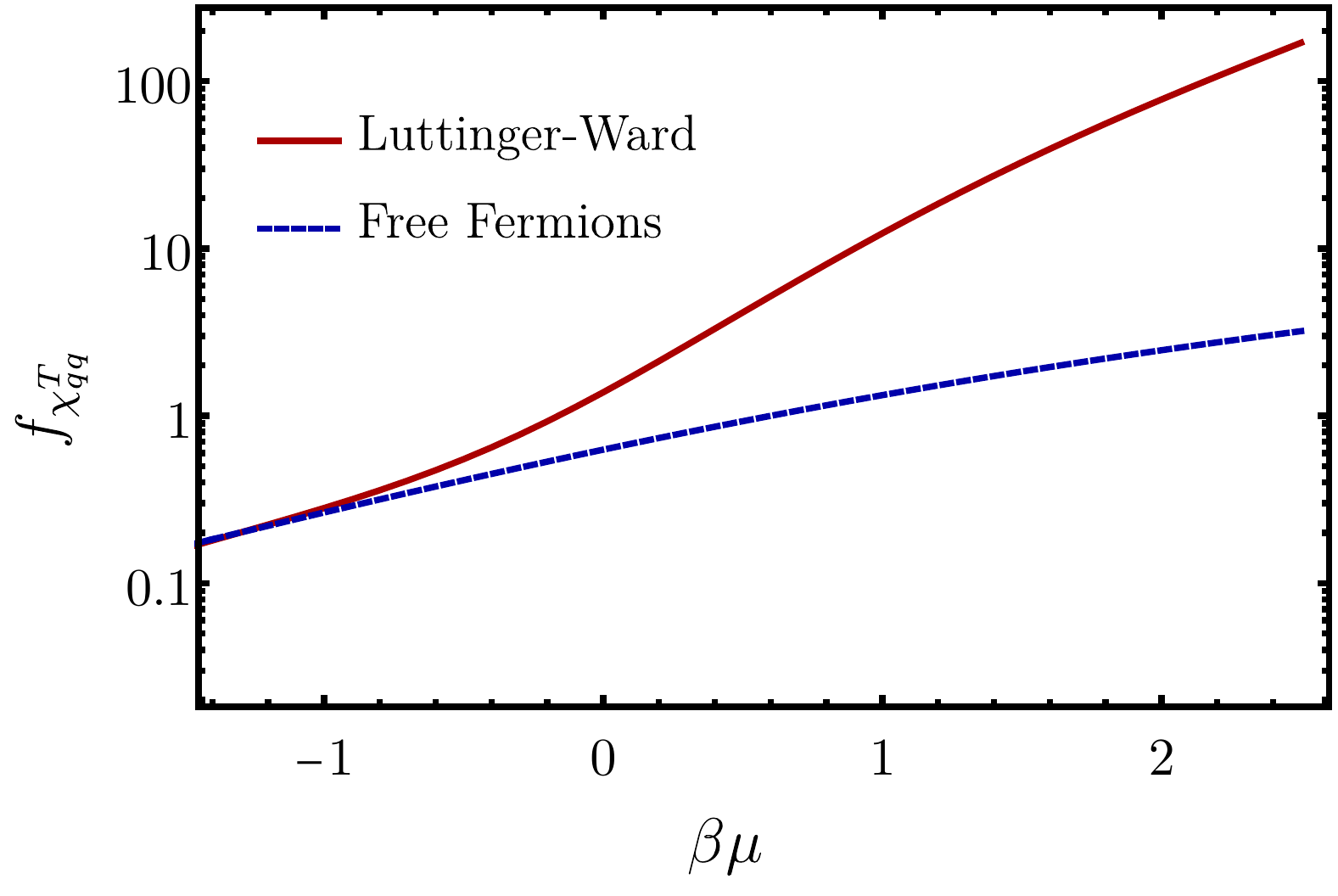}
  \caption{Thermal sum rule scaling function $f_{\chi_{qq}^T}$ vs
    $\beta\mu$ for the unitary Fermi gas: the interacting
    Luttinger-Ward result \eqref{eq:sumrule} is substantially larger
    than the free-fermion result \eqref{eq:sumrule0} in the quantum
    degenerate regime.}
  \label{fig:sumrule}
\end{figure}

After subtracting from~Eq.~\eqref{eq:sumrule} all terms that diverge
in the zero-range limit we find the exact result for the thermal
conductivity sum rule at unitarity where the pair contribution
  \eqref{eq:sumrulepair} vanishes,
\begin{align}
  \label{eq:sumrulefin}
  \chi_{qq}^T(T, \mu)
  = \chi_{qq}^{T,\text F}(T, \mu)
  \equiv \frac{T^3}{ \hbar^2\lambda_T} f_{\chi_{qq}^T}(\beta \mu)\, .
\end{align}
This defines the dimensionless quantum critical scaling function
$f_{\chi^T_{qq}}(\beta \mu)$.  In the high-temperature regime the sum
rule is given analytically by the result
\begin{multline}\label{eq:sumrule0}
  \chi_{qq}^{T(0)}(T, \mu) = -\frac{1}{4 \pi} \frac{T^3}{\hbar^2\lambda_T }
  \bigl[ 35 \Li_{7/2} (-e^{\beta \mu}) \\
  -10 (\beta \mu + \tilde{s}^{(0)} )\Li_{5/2}(-e^{\beta \mu})\bigr]
\end{multline}
of a free Fermi gas.  The entropy per particle is
$\tilde{s}^{(0)} = 5\Li_{5/2}(-e^{\beta \mu})/[2 \Li_{3/2}(-e^{\beta
  \mu})] -\beta \mu$, where $\Li_s(z)$ denotes the polylogarithm.  In
terms of density, the result approaches $m\chi_{qq}^T\to(5/2)nT^2$ for
$\beta \mu \to -\infty$.  Using the Luttinger-Ward thermodynamic data
allows us to extend the sum rule from the high-temperature regime into
the quantum degenerate regime down to the critical temperature of the
superfluid transition at $(\beta \mu) \simeq 2.5$~\cite{ku2012}.  To
obtain reliable results after the subtraction of the nonintegrable,
cutoff-dependent tails requires a precise calculation of the
self-consistent Green's and vertex functions. This has been
accomplished by using a logarithmic Fourier
transform~\cite{haines_LFT,lang_LFT} for the transformation between
real and momentum space, while the imaginary time to Matsubara
frequency transformation is performed by a discrete Fourier transform
in combination with a spline interpolation up to fifth
order~\cite{frankdiss}.  Consistency checks on thermodynamic variables
such as pressure, which contain integrals over the momentum tails as
in Eq.~\eqref{eq:sumruleF} that must remain finite in the zero-range
limit $\Lambda \to \infty$, show relative numerical errors of at most
$10^{-3}$ in the regime considered here.  This level of accuracy is a
crucial prerequisite for dealing with the more complicated asymptotics
encountered in the thermal sum rule by subtraction of the known
high-momentum behavior of the Green and vertex functions, as discussed
in detail in Appendix~\ref{sec:UV-LW}. The result for $\chi_{qq}^{T}$
is shown in Fig.~\ref{fig:sumrule}: while it agrees with
$\chi_{qq}^{T(0)}$ in the virial limit, it shows large deviations in
the quantum degenerate regime from strong pairing fluctuations, which
lead to an enhancement of up to two orders of magnitude close to the
superfluid transition.


\section{Quantum critical scattering times}
\label{sec:scatteringtimes}

Within kinetic theory, the thermal scattering time $\tau_\kappa$ is
obtained as the collision time in the Boltzmann equation in response
to temperature gradients~\cite{smith1989, schaefer2014second}. While
this approach becomes exact in the virial limit of a nondegenerate,
extremely dilute gas, it can be formally justified in the QCR only
within a large-N expansion in the number of Fermion flavors, thereby
restoring a quasiparticle picture~\cite{enss2012crit}. Quite
surprisingly, however, this method not only yields a qualitative
estimate on the physical behavior of the scattering times in the QCR,
it also allows to find a simple but quantitative approximation as we
argue below. In the Boltzmann equation the collision integral $I[f]$
is evaluated for a generic distribution function $f_p$, which deviates
from the thermal equilibrium distribution $f_p^0$ as
$f_p = f_p^0 + \delta f_p$.  For small variations $\delta f_p$, the
collision integral can be linearized as
$I[f_p] \approx H[f_p^0] \delta f_p$, where the linearized collision
operator $H[f_p^0]$ acts on $\delta f_p$ but itself only depends on
the equilibrium distribution $f_p^0$.  The solution
$\delta f_p = f_p^0 (1-f_p^0) U_p$ of the Boltzmann equation
\emph{minimizes} the scattering rate, hence the particles choose a
distribution $U_p$ to best avoid scattering.  Within a family of trial
functions $U_p$, an upper bound to the true scattering rate is found
in variational kinetic theory as \cite{smith1989, massignan2005} (for
further details see also App.~\ref{sec:var})
\begin{align}
  \label{eq:varkin}
  \tau^{-1} = \min_{U_p} \frac{(U,HU)(X,X)}{(U,X)^2}.
\end{align}
The scalar products $(A,B)=\int d\Gamma_p f_p^0 (1-f_p^0) A_p B_p$ are
defined with respect to the equilibrium distribution function $f_p^0$.
The system is driven out of equilibrium by the perturbation $X_p$: it
determines which transport channel is considered, e.g.,
$X_p=\frac{p_z}m(\varepsilon_p -(T \tilde s + \mu))$ for thermal and
$X_p=\frac{p_xp_y}m$ for shear transport.  The variational functions
$U_p$ are arbitrary functions of momentum that have the same angular
dependence as the perturbation $X_p$.

The linear collision operator $H[f^0]$ for $2\to2$ scattering between
fermionic (quasi)particles
($\vec p_1,\vec p_2\mapsto \vec p_{1'},\vec p_{2'}$) reads
\begin{multline}
  \label{eq:lincoll}
  H[f_1^0] = \int d\Gamma_2\, d\Omega\,
  \frac{d\sigma[f^0]}{d\Omega} \abs{\vec v_1-\vec v_2}
  f_2^0 (1-f_{1'}^0) (1-f_{2'}^0),
\end{multline}
where momentum conservation
$\vec p_1+\vec p_2=\vec p_{1'}+\vec p_{2'}$ and energy conservation
$\varepsilon_{p_1}+\varepsilon_{p_2} =
\varepsilon_{p_{1'}}+\varepsilon_{p_{2'}}$ are satisfied in elastic
scattering, and $\Omega$ denotes the angle between the incoming and
outgoing scattering planes.  The scattering cross section is given as
$d\sigma/d\Omega = |\tilde f|^2$ in terms of the $s$-wave scattering
amplitude $\tilde f$; in the strongly interacting Fermi gas, the
\emph{medium} scattering amplitude reads \cite{enss2012crit,
  frankdiss}
\begin{multline}
  \label{eq:scattamp}
  -\tilde f^{-1}
  = \frac1a + i\frac{\abs{\vec p_1-\vec p_2}}2 \\
  + \int d\Gamma_p\,
  \frac{2f_p^0}{\varepsilon_{p_1}+\varepsilon_{p_2}-\varepsilon_p-
    \varepsilon_{\vec p_1+\vec p_2-\vec p}+i0}
\end{multline}
to leading order in the systematic large-$N$ expansion.  While the
first two terms reproduce the $s$-wave scattering amplitude at the
two-particle level, the integral takes corrections caused by the
presence of a finite density medium into account.  At unitarity
$1/a=0$ the constant offset vanishes, and the dimensionless scattering
amplitude $\tilde f/\lambda_T$ depends on $\beta\mu$ alone.

The properties at high temperature are obtained to leading order in
the virial expansion in small fugacity $z=e^{\beta\mu}\ll1$, with
$f_p^0=e^{-\beta(\varepsilon_p-\mu)}$ the Boltzmann distribution.  The
resulting scattering times are
\begin{align}\label{eq:taukappaB}
  \frac{\tau_\kappa T}{\hbar} = \frac{45\pi}{32\sqrt2} e^{-\beta\mu}
\end{align}
and $\tau_\eta T/\hbar = \frac{15\pi}{16\sqrt2} e^{-\beta\mu}$
\cite{massignan2005} already from the first variational basis
function $U_p\propto X_p$, and corrections from higher basis functions
are less than $1.5\%$ for the shear viscosity \cite{bruun2007}.  Note
that the high-temperature results at unitarity already satisfy the
quantum critical scaling form $\tau_x T/\hbar=f_x(\beta \mu)$ with
$x=\kappa,\eta$.

In the quantum degenerate regime, one instead has to use the
Fermi-Dirac distribution
$f_p^0 = [e^{\beta(\varepsilon_p-\mu)}+1]^{-1}$.  Two competing
effects thus modify the scattering times $\tau_x$: Pauli blocking in
Eq.~\eqref{eq:lincoll} reduces the phase space for scattering and
strongly increases the scattering time, while medium scattering in
Eq.~\eqref{eq:scattamp} has the opposite effect and reduces the
scattering time.  In case of the shear viscosity the scattering cross
section $d\sigma/d\Omega$ even diverges at the superfluid transition
due to gapless pairing fluctuations if only a single variational basis
function $U_p\propto X_p$ is considered.  This would lead to the
unphysical result $\eta\to0$ at $T_c$, which arises from the
divergence of the T-matrix $\Gamma\sim Q^{-2}$ for small energies.
However, an improved variational solution in a larger basis set yields
a finite result, as is expected for the viscosity near the superfluid
transition \cite{hohenberg1977}.

\begin{figure}[t]
  \centering
  \includegraphics[width=\linewidth]{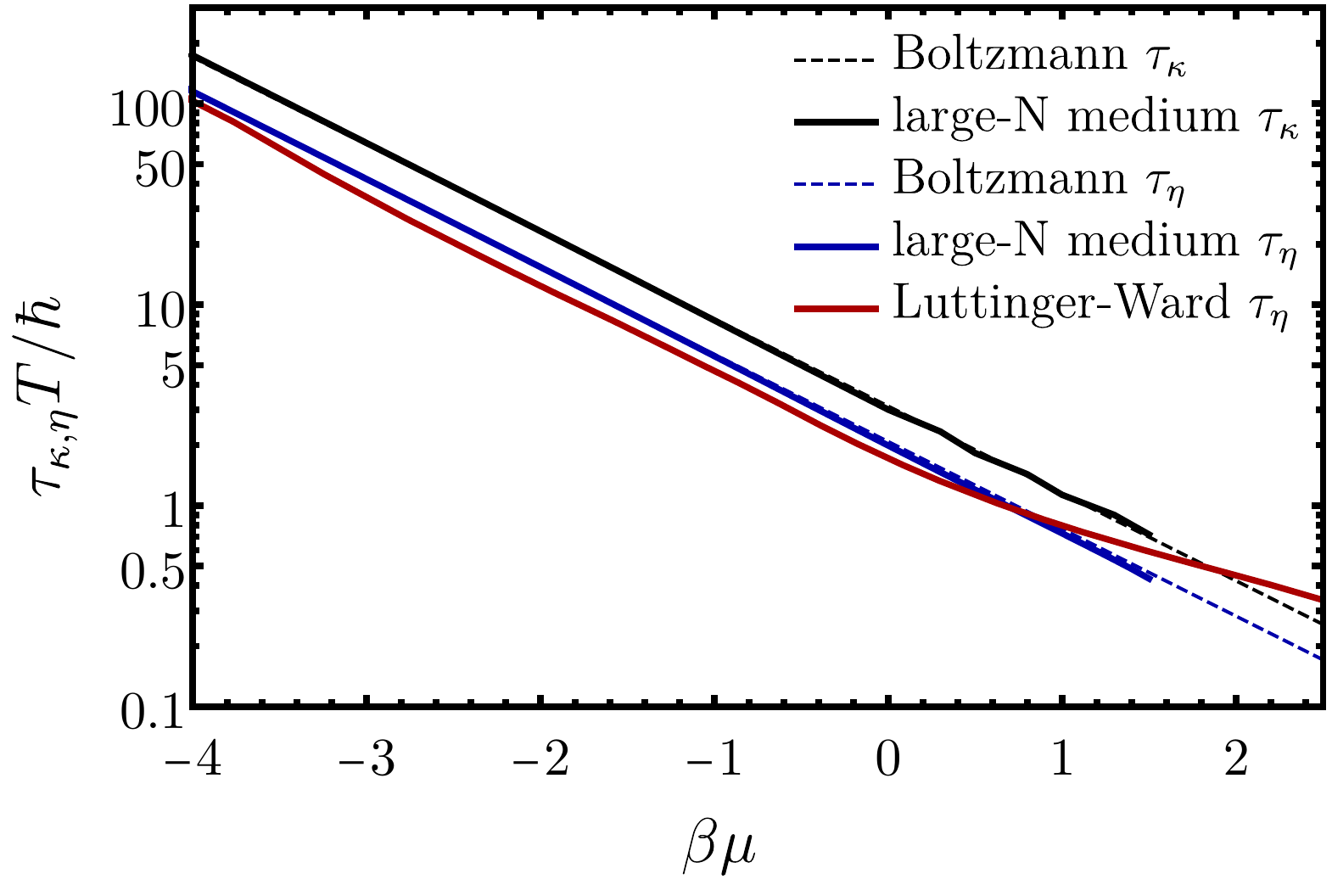}
  \caption{Thermal and viscous scattering times
    $\tau_{\kappa,\eta} T/\hbar$ vs $\beta\mu$ in the quantum critical
    region of the unitary Fermi gas.  In this regime, both Boltzmann
    and large-$N$ calculations give nearly identical results.
    Furthermore, the large-$N$ viscous scattering time (blue) agrees
    well even with the strong-coupling Luttinger-Ward computation
    \cite{enss2011} (red).}
  \label{fig:tau}
\end{figure}
The full results are shown in Fig.~\ref{fig:tau}: the surprising and
remarkable observation is that the scattering time $f_x(\beta\mu)$ is
nearly the same for the Boltzmann distribution (``Boltzmann'') and for
the Fermi-Dirac distribution (``large-N medium''), not only for
viscous~\cite{bruun2009} but also for thermal transport.  Changing the
distribution $f_p^0$ from Boltzmann to Fermi-Dirac modifies the
calculation in three places: $(i)$ in the scalar product in the
variational expression \eqref{eq:varkin}, $(ii)$ in the occupation
numbers of the collision integral \eqref{eq:lincoll}, and finally
$(iii)$ in the medium scattering amplitude \eqref{eq:scattamp}.  In
the quantum critical regime, the subtle interplay between these
effects leads to an almost perfect cancellation between the Pauli
blocking and medium scattering corrections in the large-$N$ medium
result.  We find a similar coincidence also for spin diffusion (see
App.~\ref{sec:var}).  Hence, there appears to be a general mechanism
at work that does not depend on the angular, spin or energy weight of
the driving term $X_p$.

What has not been appreciated before is that, even more remarkably,
also the strong-coupling Luttinger-Ward computations \cite{enss2011}
(red) confirm this result for the scattering time \emph{as a
  function of $\beta\mu$} for the whole quantum critical regime
$\beta\mu\lesssim1$ ($T\gtrsim2T_c$) within a $15\%$ error bound,
where the scattering time has been extracted from the relation
$\eta = p \tau_\eta$ in analogy to Eq.~\eqref{eq:kappa_def}.  We thus
conjecture that the large-$N$ expansion is similarly accurate for the
thermal scattering time $\tau_\kappa$ in the quantum critical regime,
and we use the large-$N$ result~\eqref{eq:taukappaB} henceforth.
Closer to the phase transition, however, the quantum critical scaling
crosses over into the classical critical scaling of the 3D XY
universality class near the superfluid phase transition (see
Sec.~\ref{sec:QCR} above).

At unitarity, the scattering times thus satisfy the quantum critical
scaling form \cite{sachdev2011, enss2012crit}
$\tau_x = f_x(\beta\mu) (\hbar/T)$, where the dimensionless scaling
function $f_x(\beta\mu)$ depends only on the value of $\beta\mu$, not
only in the quantum critical regime but also in the high-temperature
nondegenerate gas; in the quantum degenerate region $\beta\mu\geq0$,
the scaling function attains values of order unity
(Fig.~\ref{fig:tau}).  For spin transport (App.~\ref{sec:var}), this
is consistent with the experimental observation of quantum critical
spin drag \cite{sommer2011a} and Planckian dissipation for spin
\cite{trotzky2015, valtolina2017, enss2019spin}.


\section{Results and Quantum critical transport ratios}
\label{sec:results}

\begin{figure}[t]
  \centering
  \includegraphics[width=\linewidth]{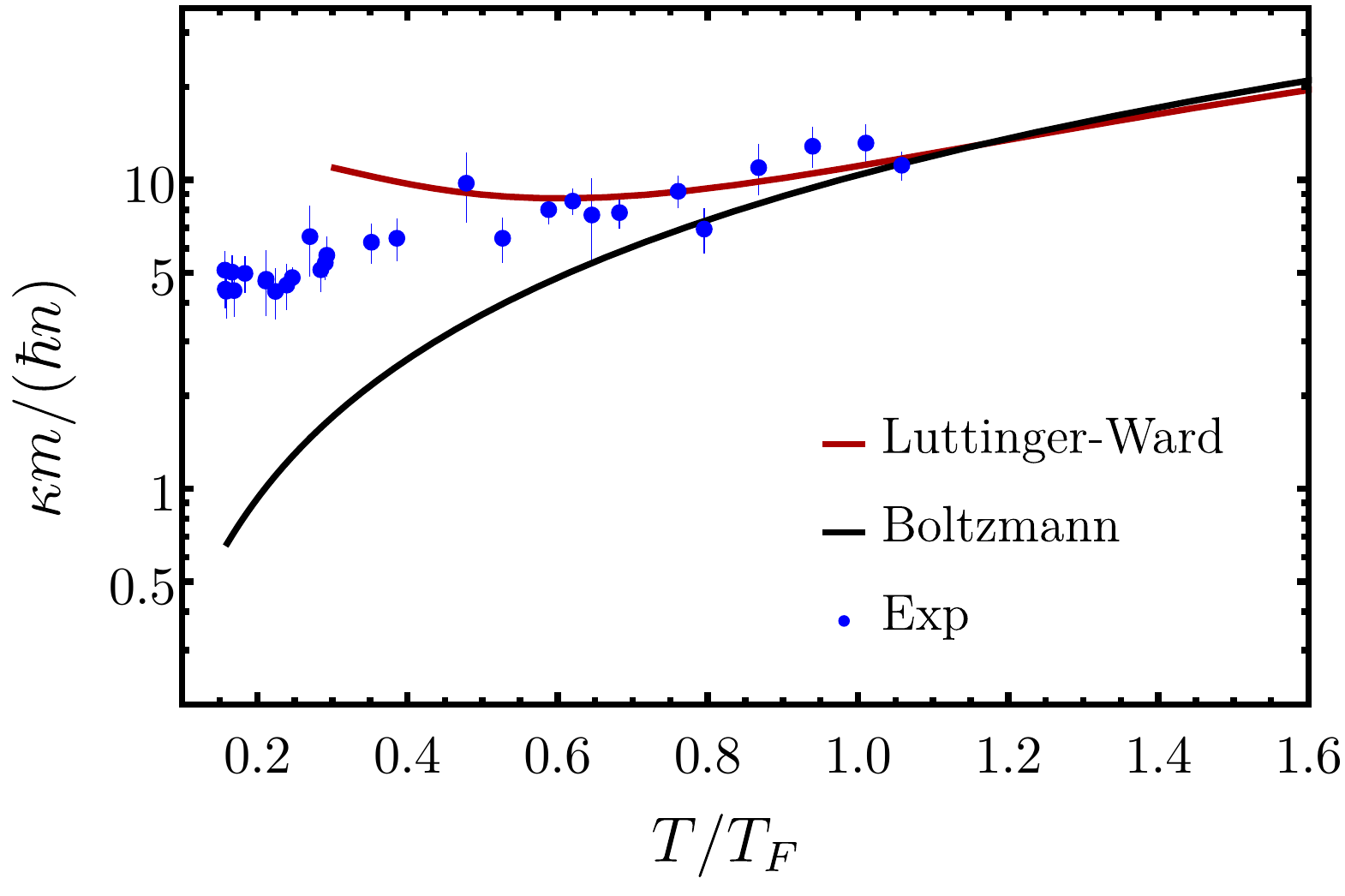}
  \caption{Thermal conductivity $\kappa$ vs temperature $T/T_F$ for
    the unitary Fermi gas from Luttinger-Ward calculations (red line)
    and from experiment \cite{patel2019}.  $\kappa$ saturates in
      the quantum degenerate regime and exhibits a shallow minimum of
      $\kappa/n \approx 8.7\hbar/m$ at $T/T_F\approx0.6$.}
  \label{fig:kappa}
\end{figure}
Based on the hydrodynamic arguments from above, we arrive at the first
prediction for the thermal conductivity \eqref{eq:kappa_def},
$\kappa T=\chi_{qq}^T\tau_\kappa$, in the quantum critical regime, as
shown in Fig.~\ref{fig:kappa}.  Here, $\chi_{qq}^T$ is evaluated
within Luttinger-Ward theory (Fig.~\ref{fig:sumrule}) and combined
with the thermal scattering time in the Boltzmann
limit~\eqref{eq:taukappaB}.  In the limit $\beta \mu \to -\infty$ one
finds the Boltzmann value for the thermal conductivity
$\kappa^B= 225/(64 \sqrt{2}) \, T/(\hbar \lambda_T)$ \cite{braby2010}
by using the noninteracting sum rule~\eqref{eq:sumrule0}.  At lower
temperatures, however, the strong enhancement of the thermal sum rule
implies a significantly larger thermal conductivity as opposed to the
result from the Boltzmann equation (cf.\ Fig.~\ref{fig:sumrule}).

\begin{figure}[t]
  \centering
  \includegraphics[width=\linewidth]{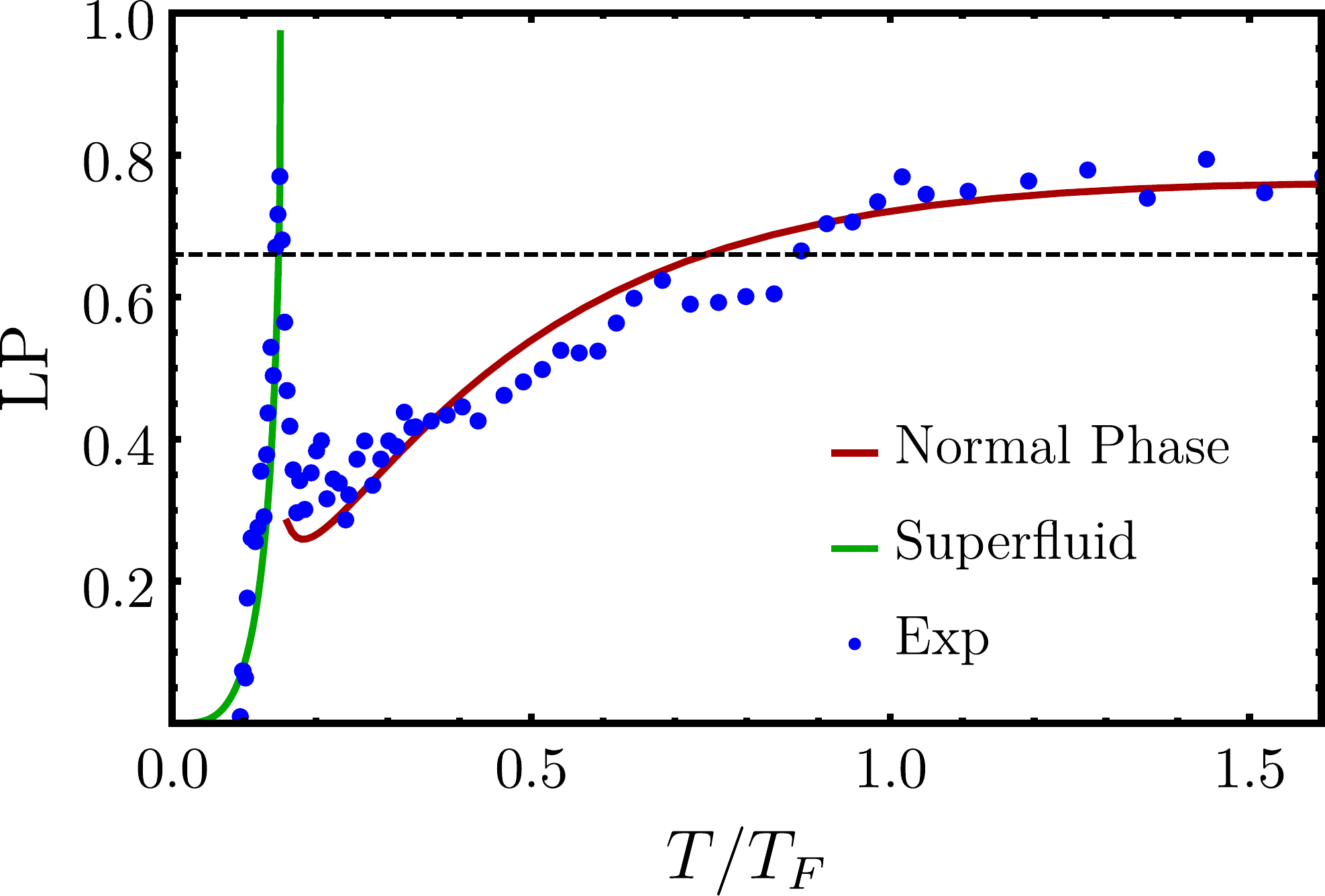}
  \caption{Landau-Placzek ratio $\text{LP}=c_p/c_V-1$ vs temperature
    $T/T_F$ from Luttinger-Ward calculations (green below $T_c$
    \cite{haussmann2007}, red above $T_c$ \cite{frankdiss}) and from
    experiment \cite{ku2012}.  The dashed line indicates the
    high-temperature limit $\text{LP}=2/3$.}
  \label{fig:LP}
\end{figure}
Weighting the thermal diffusion $D_T$ with the thermodynamic
Landau-Placzek ratio $\text{LP}=c_p/c_V-1$ (Fig.~\ref{fig:LP}) yields
the thermal contribution $\text{LP}\times D_T$ to the sound diffusion
$D_\text{sound}$ shown in Fig.~\ref{fig:diffusion}.  At low
temperatures above $T_c$ the decrease of $\text{LP}$ seems to suggest
that heat diffusion becomes less important for sound attenuation near
$T_c$, but this is more than compensated by the increase of $D_T$
which makes heat diffusion rather more important.
\begin{figure}[t]
  \centering
  \includegraphics[width=\linewidth]{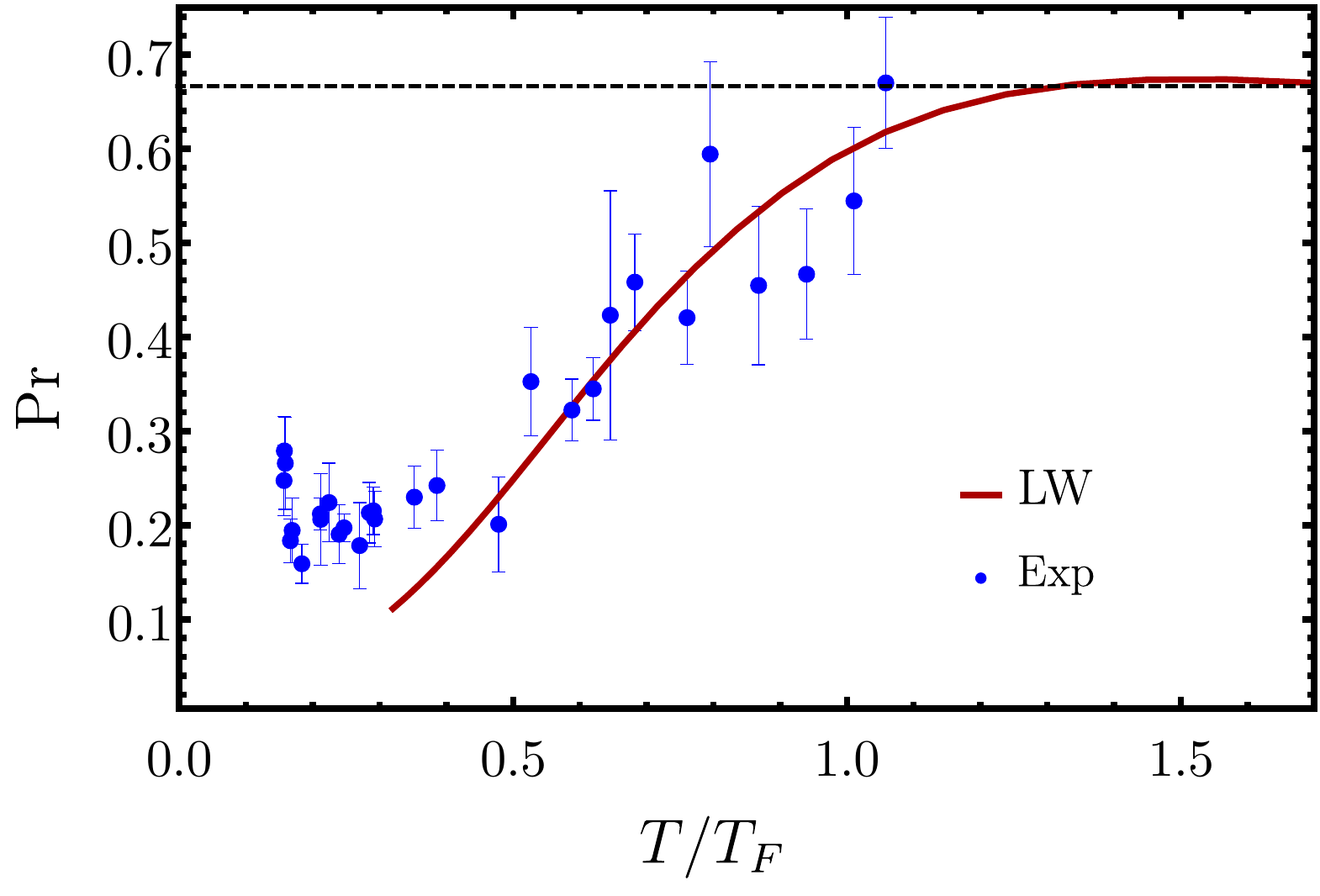}
  \caption{Prandtl number $\text{Pr}=D_\eta/D_T$ vs temperature
    $T/T_F$ from Luttinger-Ward calculations (red line) and sound
    attenuation measurements \cite{patel2019}; the dashed line marks
    the high-temperature limit $\text{Pr}=2/3$.}
  \label{fig:prandtl}
\end{figure}

A fluid is characterized by the relative importance of different
transport channels, which is quantified by transport ratios.  Here, we
consider the Prandtl number, which is defined as the ratio of shear
and thermal diffusivities ($D_T$ is reported in
Fig.~\ref{fig:diffusion}),
\begin{align}
  \text{Pr} = \frac{D_\eta}{D_T}
  = \frac{(p/mn)\tau_\eta}{(\chi_{qq}^T/c_pT)\tau_\kappa}
  = \frac{pc_pT}{mn\chi_{qq}^T}\times \frac{\tau_\eta}{\tau_\kappa}.
\end{align}
As the last term shows, the transport ratio is a product of a
thermodynamic term that incorporates nontrivial temperature scaling
from the full equation of state, and a ratio of transport times which
we have found to remain nearly constant at $\tau_\eta/\tau_\kappa=2/3$
throughout the quantum critical regime.  Therefore, in the unitary gas
the transport ratios derive their temperature dependence predominantly
from the equation of state, and we use the best available
Luttinger-Ward equation of state \cite{haussmann2007, frankdiss} to
obtain the theory prediction for the Prandtl number in
Fig.~\ref{fig:prandtl}.  Note that $\text{Pr}$ starts from a value of
$2/3$ in the high-temperature limit and then grows to about $0.7$ near
$T\approx T_F$, before it falls to much smaller values below $0.2$
near the superfluid transition.  This nonmonotonic behavior results
from the Landau-Placzek ratio \cite{frankdiss, hu2018} shown in
Fig.~\ref{fig:LP} for the unitary Fermi gas, and is consistent with
the virial expansion \cite{braby2010}.  At the classical superfluid
phase transition (Model F) \cite{hohenberg1977} one expects that
$\eta$ remains finite while $\kappa$ diverges according to
Eq.~\eqref{eq:critical}, suggesting a vanishing $\text{Pr}\to0$.  This
nonmonotonic behavior is very well confirmed by a recent measurement
of sound attenuation in the unitary gas \cite{patel2019}.  The value
of the Prandtl number also has an important interpretation in terms of
possible nonrelativistic gravity duals, which, however, predict
$\text{Pr}=1$ \cite{rangamani2009} and can therefore be excluded as a
model for the unitary Fermi gas.  Another important transport ratio is
the bulk-to-shear viscosity ratio $\zeta/\eta$ computed in
\cite{enss2019bulk}, which shows that viscous transport occurs via
quasiparticles only for $T\gtrsim T_F$ but deviates in the quantum
degenerate regime.  The Schmidt number $\text{Sc}=D_\eta/D_s$
comparing shear with spin transport is shown in
Fig.~\ref{fig:schmidt}, see Appendix~\ref{sec:var}.


\section{Discussion}
\label{sec:discussion}

In conclusion, we have found that transport scattering times
$\tau_\kappa$ and $\tau_\eta$ in the quantum critical regime follow a
remarkably simple scaling law, which extends to the vicinity of the
superfluid transition where pairing fluctuations become dominant.  We
have chosen specifically the unitary gas where the quantum critical
regime extends to high temperature to demonstrate this point.  This
information is combined with a new exact sum rule for thermal
transport, which depends on the equation of state and thermal
operators beyond, to predict the thermal conductivity $\kappa$ in the
quantum degenerate regime.  For $\kappa$ and the Prandtl number
$\text{Pr}$ we find good agreement with recent experiments
\cite{patel2019}.

The remarkable coincidence of the quantum critical scattering times
from the high-temperature Boltzmann calculation and the
strong-coupling large-$N$ and Luttinger-Ward results is a unique
feature of the quantum critical point at infinite scattering length
$1/a=0$: the scattering times must follow the quantum critical scaling
form, which in the particular case of the unitary Fermi gas must
extend up to high temperature by dimensional analysis, in contrast to
lattice models.  At high temperature, the scattering times are
reliably obtained from kinetic theory as $\tau T/\hbar\propto z^{-1}$
proportional to the inverse fugacity.  Now quantum critical scaling
predicts that this form continues from the dilute gas throughout the
QCR until near $T_c$.  This remarkable observation is supported by the
fact that it leads to good agreement with recent experimental data in
the regime where quantum critical scaling can be applied.  It will be
interesting to see if our approach can be extended to other types of
QCPs.

While at unitarity the bosonic part of the exact sum rule
\eqref{eq:sumrule} provides only a regularization, away from unitarity
($1/a\neq0$) it gives a new regular contribution that arises from
local pair fluctuations, the so-called contact correlations
\cite{enss2019bulk}.  This new contribution to thermal transport is
not captured by fermionic kinetic theory and is particularly important
at low temperatures near the superfluid phase transition.

We find that not only shear and spin diffusion, but also the thermal
diffusion $D_T$ in units of $\hbar/m$ exhibit quantum limited
diffusion near $T_c$.  For thermal transport, the diffusion minimum
$D_T\simeq 4.2\,\hbar/m$ occurs well in the quantum critical
region at $T \simeq 0.7\, T_F$ (see Fig.~\ref{fig:diffusion}).  Hence,
the quantum degenerate unitary Fermi gas is a nearly perfect fluid not
only regarding momentum transport but also for thermal transport.

With current sound propagation measurements in box traps reaching into
the superfluid regime \cite{patel2019, bohlen2020}, it will be
particularly interesting to study critical scaling of the transport
coefficients and observe the increase of $D_T$ shown in
Fig.~\ref{fig:diffusion}.  This, as well as the related monotonic
decrease of the Prandtl number indicated in Fig.~\ref{fig:prandtl}, is
due to the growing thermal conductivity associated with the crossover
to classical critical fluctuations as expressed asymptotically in
Eq.~\eqref{eq:critical}.  For the sound diffusion $D_\text{sound}(T)$
(Fig.~\ref{fig:diffusion}) both our quantum critical prediction and
the experimental data indicate a monotonic decrease, while at even
lower temperatures $T/T_F\lesssim 0.2$ an increase of $D_\text{sound}$
is again theoretically expected from critical fluctuations.  In the
strongly interacting 2D Fermi gas the recently observed quantum scale
anomaly \cite{murthy2019} will have a large effect on the transport
coefficients.

\begin{acknowledgments}
  We thank N.~Defenu and M.~W. Zwierlein for stimulating discussions,
  and M.~W. Zwierlein for sharing experimental data.  This work is
  supported by the Deutsche Forschungsgemeinschaft (DFG, German
  Research Foundation) via Collaborative Research Centre ``SFB1225''
  (ISOQUANT) and under Germany’s Excellence Strategy
  ``EXC-2181/1-390900948'' (the Heidelberg STRUCTURES Excellence
  Cluster).
\end{acknowledgments}


\appendix

\section{Energy current Ward identity}
\label{sec:ward}

Following Polyakov \cite{polyakov1969}, energy conservation leads to
the continuity equation
$\partial_t \Hat{\mathcal H} + \nabla\cdot \vec \hat {\bm\jmath}^E =
0$ between the energy density operator (Hamiltonian) and the energy
current operator defined in Eq.~\eqref{eq:currentsreal}.  The
expectation values of this operator equation together with two
additional fermion operators then lead to the Ward identity for the
full energy current vertex (cf.~\eqref{eq:vertex})
\begin{align}
\begin{split}
  & \vec q\cdot \mathcal T_\sigma^E(\vec p,\varepsilon,\vec q,\omega=0)\\
  & = G_\sigma(\vec p,\varepsilon)G_\sigma(\vec p+\vec q,\varepsilon) \vec q\cdot 
  \tilde{\mathcal T}_\sigma^E(\vec p,\varepsilon,\vec q,\omega=0) \\
  & = \Bigl(\varepsilon+\mu-\frac{\vec q\cdot (\vec p+\vec q)}{2m}\Bigr)
    G_\sigma(\vec p+\vec q,\varepsilon) \\
  & \qquad - \Bigl(\varepsilon+\mu+\frac{\vec q\cdot \vec p}{2m}\Bigr)
    G_\sigma(\vec p,\varepsilon) \\
  & = \vec q\cdot \frac{\partial G_\sigma[\mathbf h]}{\partial \mathbf h_q}.
  \end{split}
\end{align}
The amputated energy current vertex is then
\begin{align}
  \begin{split}
  & \vec q\cdot \tilde{\mathcal T}_\sigma^E(\vec p,\varepsilon,\vec q,\omega=0)\\
  & = \Bigl(\varepsilon+\mu-\frac{\vec q\cdot (\vec p+\vec q)}{2m}\Bigr)
    G_\sigma^{-1}(\vec p,\varepsilon) \\
  & \qquad - \Bigl(\varepsilon+\mu+\frac{\vec q\cdot \vec p}{2m}\Bigr)
    G_\sigma^{-1}(\vec p+\vec q,\varepsilon) \\
  & = -\vec q\cdot \frac{\partial G_\sigma^{-1}[\mathbf h]}{\partial
    \mathbf h_q},
  \end{split}
\end{align}
and by inserting the noninteracting Green functions $G_0^{-1}(\vec
p,\varepsilon) = \varepsilon - \varepsilon_p + \mu$ one finds the
matrix elements of the bare energy current \emph{operator}
(cf.~\eqref{eq:currents}),
\begin{align}
  \tilde{\mathcal T}_\sigma^{E(0)}(\vec p,\vec q)
  = j_\sigma^E(\vec p,\vec q)
  = \frac{\vec p+\vec q/2}m \times \frac{\vec p\cdot (\vec p+\vec q)}{2m}.
\end{align}

In the two-channel model formulated in terms of both fermions and
pairs, we have to introduce a new \emph{bosonic} Ward identity beyond
the one given by Polyakov \cite{polyakov1969}.  We find for the full
bosonic energy current vertex (cf.~\eqref{eq:vertex})
\begin{align}
\begin{split}
  & \vec q\cdot \mathcal T_\text{pair}^E(\vec Q,\Omega,\vec q,\omega=0)\\
  & = \Gamma(\vec Q,\Omega)\Gamma(\vec Q+\vec q,\Omega) \vec q\cdot 
  \tilde{\mathcal T}_\text{pair}^E(\vec Q,\Omega,\vec q,\omega=0) \\
  & = \Bigl(\Omega+2\mu+\frac{\vec q\cdot \vec Q}{2m}\Bigr)
    \Gamma(\vec Q+\vec q,\Omega) \\
  & \qquad - \Bigl(\Omega+2\mu-\frac{\vec q\cdot (\vec Q+\vec q)}{2m}\Bigr)
    \Gamma(\vec Q,\Omega) \\
  & = \vec q\cdot \frac{\partial \Gamma[\mathbf h]}{\partial \mathbf h_q}.
  \end{split}
\end{align}
The amputated bosonic energy current vertex is then given by
\begin{align}
\begin{split}
  & \vec q\cdot\tilde{\mathcal T}_\text{pair}^E(\vec Q,\Omega,\vec q,\omega=0) \\
  & = \Bigl(\Omega+2\mu+\frac{\vec q\cdot \vec Q}{2m}\Bigr)
    \Gamma^{-1}(\vec Q,\Omega) \\
  & \qquad - \Bigl(\Omega+2\mu-\frac{\vec q\cdot (\vec Q+\vec q)}{2m}\Bigr)
    \Gamma^{-1}(\vec Q+\vec q,\Omega) \\
  & = -\vec q\cdot \frac{\partial \Gamma^{-1}[\mathbf h]}{\partial
    \mathbf h_q}. 
  \end{split}
\end{align}
When inserting the bare bosonic Green function $\Gamma_0(\vec
Q,\Omega) = \bar g(\Lambda)$, one recovers the matrix elements of the
bosonic energy current operator (cf.~\eqref{eq:currents}),
\begin{align}
  \tilde{\mathcal T}_\text{pair}^{E(0)}(\vec Q,\vec q)
  = j^E_\text{pair}(\vec Q,\vec q)
  = \frac{\vec Q+\vec q/2}{m\bar g(\Lambda)}.
\end{align}
In the limit $\omega=0$, $\vec q\to0$ the homogeneous Ward identities
and current operators result in the expressions that are given in the
main text:
\begin{align*}
  \mathcal T_\sigma^E(\vec p,\varepsilon)
  & = -\frac{\vec p}m\,G_\sigma(\vec p,\varepsilon) + (\varepsilon+\mu)\,
  \frac{\partial G_\sigma(\vec p,\varepsilon)}{\partial\vec p}, \\
  \tilde{\mathcal T}_\sigma^E(\vec p,\varepsilon)
  & = -\frac{\vec p}m\,G_\sigma^{-1}(\vec p,\varepsilon) - (\varepsilon+\mu)\,
  \frac{\partial G_\sigma^{-1}(\vec p,\varepsilon)}{\partial\vec p}, \\
  \tilde{\mathcal T}_\sigma^{E(0)}(\vec p,\varepsilon)
  & = \frac{\vec p}m\,\varepsilon_p, \\
  \mathcal T^E_\text{pair}(\vec Q,\Omega)
  & = \frac{\vec Q}m\,\Gamma(\vec Q,\Omega) +
    (\Omega+2\mu)\,\frac{\partial\Gamma(\vec Q,\Omega)}{\partial\vec Q}, \\
  \tilde{\mathcal T}^E_\text{pair}(\vec Q,\Omega)
  & = \frac{\vec Q}m\,\Gamma^{-1}(\vec Q,\Omega) -
    (\Omega+2\mu)\,\frac{\partial\Gamma^{-1}(\vec Q,\Omega)}{\partial\vec Q}, \\
  \tilde{\mathcal T}^{E(0)}_\text{pair}(\vec Q,\Omega)
  & = \frac{\vec Q}m\,\frac{1}{\bar g(\Lambda)}.
\end{align*}

\section{UV asymptotics of correlation functions}
\label{sec:UV-LW}

The power-law tails typical for correlation functions in the
zero-range limit arise from interaction effects, which are encoded in
the fermionic self-energy $\Sigma(\mathbf p \to \infty, \tau \to 0^-)$
and in the pair propagator
$\Gamma(\mathbf Q \to \infty, \tau \to 0^-)$. In order to formulate
the Luttinger-Ward theory it is more convenient to make use of the
(anti-)periodicity of (fermionic) bosonic correlation functions in
imaginary time and to consider the limit $\tau \to \beta^-$ instead of
$\tau \to 0^-$.  Within the self-consistent T-matrix approximation for
the Luttinger-Ward grand potential the self-energy of unpolarized
fermions is given by (henceforth $\hbar\equiv1$)
\begin{align}\label{eq:se1}
  \Sigma(\mathbf p, \tau) = -\int \frac{d^3 Q}{(2 \pi)^3}
  \Gamma(\mathbf Q, \tau) G(\mathbf Q - \mathbf p, \beta-\tau)\, .
\end{align}  
In the limit of large momenta both the Green's and the vertex
functions approach their vacuum forms~\cite{vanhoucke2019BDMC}
\begin{align}
\begin{split}
  G(\mathbf p\to \infty ,\tau) & \to G_v(\mathbf p ,\tau)
  \simeq - e^{- \tau \varepsilon_p} \\
  \Gamma(\mathbf Q \to \infty ,\tau) & \to \Gamma_v(\mathbf Q,\tau)
  \simeq - \frac{4 \sqrt{\pi}}{m^{3/2}\sqrt{\tau}}
  e^{- \tau \varepsilon_{Q}/2} \, .
\end{split}
\end{align} 
Moreover, in the vacuum limit all diagrams vanish except for the
particle-particle ladders which represent the exact T-matrix for
two-particle scattering in quantum mechanics. Therefore, the
Luttinger-Ward approach includes the correct exponents of the momentum
tails. For $\tau \to \beta^-$ and $\mathbf p \to \infty$ we retain
only the dominant contributions to the momentum integral which arise
from the regions $|\mathbf{Q}| \ll |\mathbf{p}|$ and
$|\mathbf{Q} -\mathbf{p}| \ll |\mathbf{p}|$. This allows one to expand
the Green's function in the form
\begin{widetext}
\begin{align}
  &G_v(\mathbf{Q}-\mathbf{p}, (\beta-\tau)\to 0^+) \simeq
  - e^{-(\beta - \tau)\varepsilon_p} 
      \Bigl[1 +(\beta- \tau)\Bigl(\frac{\mathbf Q \cdot \mathbf p}{m}
      -\varepsilon_Q\Bigr) + \frac{(\beta-\tau)^2}{2} \frac{(\mathbf{Q \cdot \mathbf p})^2}{m^2}
    + \dotsm\Bigr]\, ,
\end{align}
and we obtain for the self-energy
\begin{align}\label{eq:SE-conv}
\begin{split}
  & \Sigma(\mathbf p \to \infty , \tau \to \beta^-)  \\ 
  &\simeq e^{-(\beta - \tau) \varepsilon_p}\Bigl[ 
    \Gamma(\mathbf x= \mathbf 0, \tau \to \beta^-) + \left(\frac{4 (\beta-\tau)^2}{3}\varepsilon_p
      - (\beta -\tau) \right)
    \int \frac{d^3Q}{(2\pi)^3} \varepsilon_Q \Gamma(\mathbf{Q} ,\tau \to \beta^-) \Bigr] \\
  & = e^{-(\beta - \tau) \varepsilon_p}\Bigl[ 
    \Gamma(\mathbf x=\mathbf 0, \tau \to \beta^-)+ \left(\frac{4 (\beta-\tau)^2}{3}\varepsilon_p
      - (\beta -\tau) \right) \left(- \frac{\nabla^2}{2m} \right)
    \Gamma(\mathbf x ,\tau \to \beta^-)_{\mathbf x=\mathbf 0} \Bigr] \, .
\end{split}
\end{align}
Here we have assumed that the momentum integral of the pair propagator
is finite, which we show below in Eq.~\eqref{eq:Gamma-asy}.  From our
Luttinger-Ward data we find the behavior
\begin{align}\label{eq:coeffGamma}
  -\Gamma(\mathbf x \to \mathbf 0, \tau \to \beta^-) 
  = \mathcal C/m^2 
  +\Gamma_1 \cdot (\beta- \tau) + \Gamma_{3/2}\cdot (\beta- \tau)^{3/2}
  + \Gamma_{x^2}\cdot x^2 +\dotsm \, ,
\end{align}
where $\Gamma_{1,3/2,x^2}$ denote new coefficients while the leading
order is determined by the Tan contact $\mathcal C$ according to
Eq.~\eqref{eq:contact}.  The anomalous power $(\beta-\tau)^{3/2}$ is
generated by the self-consistent iteration but unbiased by the
necessary analytic subtractions~\cite{frankdiss} which consider only
contributions to the limit $\tau \to 0^+$.  This result implies for
the self-energy
\begin{align}\label{eq:SEMatsu}
\begin{split}
  \Sigma(\mathbf p \to \infty , \tau \to \beta^-)
  & \to 
  -e^{-(\beta - \tau) \varepsilon_p} \Bigl[ \mathcal C + \Gamma_1 (\beta -\tau)
  + \Gamma_{3/2} (\beta- \tau)^{3/2} - \frac{\Gamma_{x^2}}{m}
    \left(4(\beta-\tau)^2\varepsilon_p - 3(\beta -\tau) \right)  \Bigr] \, , \\
  \Sigma(\mathbf p \to \infty , \epsilon_n)
  & \to  \frac{\mathcal C/m^2}{i \epsilon_n  +\varepsilon_p}
  + \frac{\Gamma_1+ \frac{3}{m} \Gamma_{x^2}}{(i \epsilon_n + \varepsilon_p)^2 }
  -\frac{\frac{8}{m}\Gamma_{x^2} \varepsilon_p}{(i \epsilon_n + \varepsilon_p)^3}
  + \frac{3 \sqrt{\pi} \Gamma_{3/2} }{4 (i \epsilon_n +\varepsilon_p)^{5/2}} \, .
\end{split}
\end{align}
\end{widetext}
Here, the second line is obtained from the first one by Fourier
transform to Matsubara frequencies.  Using the Dyson equation
$G^{-1}(\mathbf{p},\epsilon_n) =
G^{-1}_{0}(\mathbf{p},\epsilon_n)-\Sigma(\mathbf p,
\epsilon_n) $ one can determine the asymptotic power laws of the
momentum distribution
\begin{align}\label{eq:momdistasymp}
\begin{split}
  n(\mathbf p \to \infty)
  & = 
  \frac1\beta \sum_n  \frac{\Sigma(\mathbf p \to \infty,\epsilon_n )}
  {(i \epsilon_n - \varepsilon_p)^2} \\
  & = \frac{\mathcal{C}}{p^4}
  +\frac{\Gamma_1-\frac{3}{m}\Gamma_{x^2}}{4 \varepsilon_p^3}
  + \frac{15 \sqrt{2 \pi} \Gamma_{3/2} }{128\varepsilon_p^{7/2}} \dotsm\, ,
 \end{split}
\end{align}
which is indeed of the form~\eqref{eq:Tan-momentum} stated in the main
text.

We turn now to the UV-behavior of the pair propagator
$\Gamma(\mathbf Q \to \infty, \tau \to \beta^- )$. In the ladder
approximation it can be expressed via the Bethe-Salpeter equation
\begin{align}
  \Gamma(\mathbf Q, \Omega_n) = \frac{1}{1/g + M_{pp}(\mathbf Q, \Omega_n)} \, ,
\end{align}
where $M_{pp}$ denotes the renormalized particle-particle bubble
\begin{align}
  M_{pp} (\mathbf Q,\tau) = \int \frac{d^3p}{(2\pi)^3}
  G (\mathbf p ,\tau)  G (\mathbf Q -\mathbf p ,\tau) \, .
\end{align}
However, the cancellation of divergent terms in the zero-range limit
affects only the behavior $\tau \to 0^+$ and needs not to be taken
further into account. Employing analogous arguments that led from the
convolution~\eqref{eq:se1} to the result~\eqref{eq:SE-conv} while
using the asymptotic form~\cite{vanhoucke2019BDMC}
$G(\mathbf p \to \infty , \tau \to \beta^-) \to - (\mathcal C/p^4)
\exp(-\varepsilon_p (\beta -\tau))$, we obtain the limiting behavior
\begin{align}
  M^\beta_{pp} (\mathbf Q, \tau )
  = M_{pp} (\mathbf Q \to \infty, \tau \to \beta^-)
  \to \frac{n \mathcal C}{Q^4} e^{-\varepsilon_Q (\beta -\tau)} \, ,
\end{align} 
where we have inserted the total density
$n = -2 G (\mathbf x=0,\beta^-)$. Transforming this to
frequency space yields
\begin{align}
  M_{pp}^\beta (\mathbf Q, \Omega_n)
  =  \frac{n\mathcal{C} }{Q^4} \frac{1}{\varepsilon_Q + i \Omega_n} \, ,
\end{align}
which combines with the leading contribution in the vacuum limit
$M_{pp}(\mathbf{Q} ,\Omega_n \to \infty) \to
-m^{3/2}\sqrt{\varepsilon_Q-2 i \Omega_n}/(2^{5/2} \pi)$ to yield the
pair momentum distribution (of dimension wavenumber due to the
anomalous dimension of the contact operator),
\begin{align}\label{eq:Gamma-asy}
  n_\text{pair}(Q)
  = -m^2 \Gamma(\mathbf{Q},\tau \to \beta^-)
  = \frac{64\pi^2}{3} \frac{n \mathcal{C}}{Q^6} + \dotsm\,.
\end{align}


\section{Variational kinetic theory}
\label{sec:var}

In this appendix we explain how to evaluate the variational bound on
the transport scattering rate \eqref{eq:varkin} in a larger basis set.
Specifically for the shear viscosity, $X_{\vec p}=p_xp_y/m$ denotes
the shear perturbation and $U_{\vec p}$ has the same quadrupole
symmetry as $X_{\vec p}$, hence it differs from $X_{\vec p}$ only by a
spherically symmetric function of $p^2$.  One can expand
$U_{\vec p}=\sum_ic_iU_i(\vec p)$ in orthogonal (but not necessarily
normalized) basis functions $U_i$ with $(U_i,U_j)=u_i\delta_{ij}$.  A
particularly convenient choice is setting $U_1(\vec p)=X(\vec p)$ and
finding orthogonal $U_i$, $i>1$, by the Gram-Schmidt method, which
simplifies Eq.~\eqref{eq:varkin} to
\begin{align}
  \label{eq:vartauinv}
  \tau^{-1} = \min_{U(p)} \frac{(U,HU)}{c_1^2(X,X)}.
\end{align}
The collision integral is normalized by $(X,X)$, which in the case of
the shear viscosity is proportional to the pressure of the ideal Fermi
gas, $(X,X)=-\lambda^{-3}_T T^2\Li_{5/2}(-e^{\beta\mu})$.  Denoting
the matrix elements of the positive linear operator $H$ as
$H_{ij}=(U_i,HU_j)$, the stationarity of $\tau^{-1}$ with respect to
variations in $U$ requires
$\tau^{-1}\delta_{i1}c_1(X,X)=\sum_jH_{ij}c_j$.  The minimum principle
for $\tau^{-1}$ then implies that each minimization within a finite
subspace of $U_i$ for $i=1,\dotsc,M$ provides an upper bound on the
true value of $\tau^{-1}$, which can be successively improved
(lowered) by increasing $M$.  Equivalently, this can be expressed as a
lower bound on the scattering time,
\begin{align}
  \label{eq:vartau}
  \tau \geq (H^{-1})_{11}(X,X),
\end{align}
in terms of the $(1,1)$ element of the inverse matrix of $H_{ij}$.
Results for the viscous scattering time $\tau_\eta$ in the unitary
Fermi gas are shown in Fig.~\ref{fig:tau} in the main text.  The
surprising observation for the viscous scattering time $\tau_\eta$ at
unitarity is that it has nearly the same value both with a Boltzmann
distribution and with a Fermi-Dirac distribution, but only if a full
variational basis set beyond the first basis function $U_1$ is used.

Analogously, a similar observation is made for the heat conductivity
with driving term $X_{\vec p}=(\varepsilon_p-w)\frac{\vec p}m$.
Again, we choose a set of variational basis functions
$U_i(\vec p)=p^{2(i-1)}X_{\vec p}$ for $i=1,\dotsc,M$ \cite{frankdiss}
and find that the thermal transport scattering time converges rapidly
already with the first three basis functions, but differs markedly
from the result with only the first basis function $U_1$.

\begin{figure}[t]
  \centering
  \includegraphics[width=\linewidth]{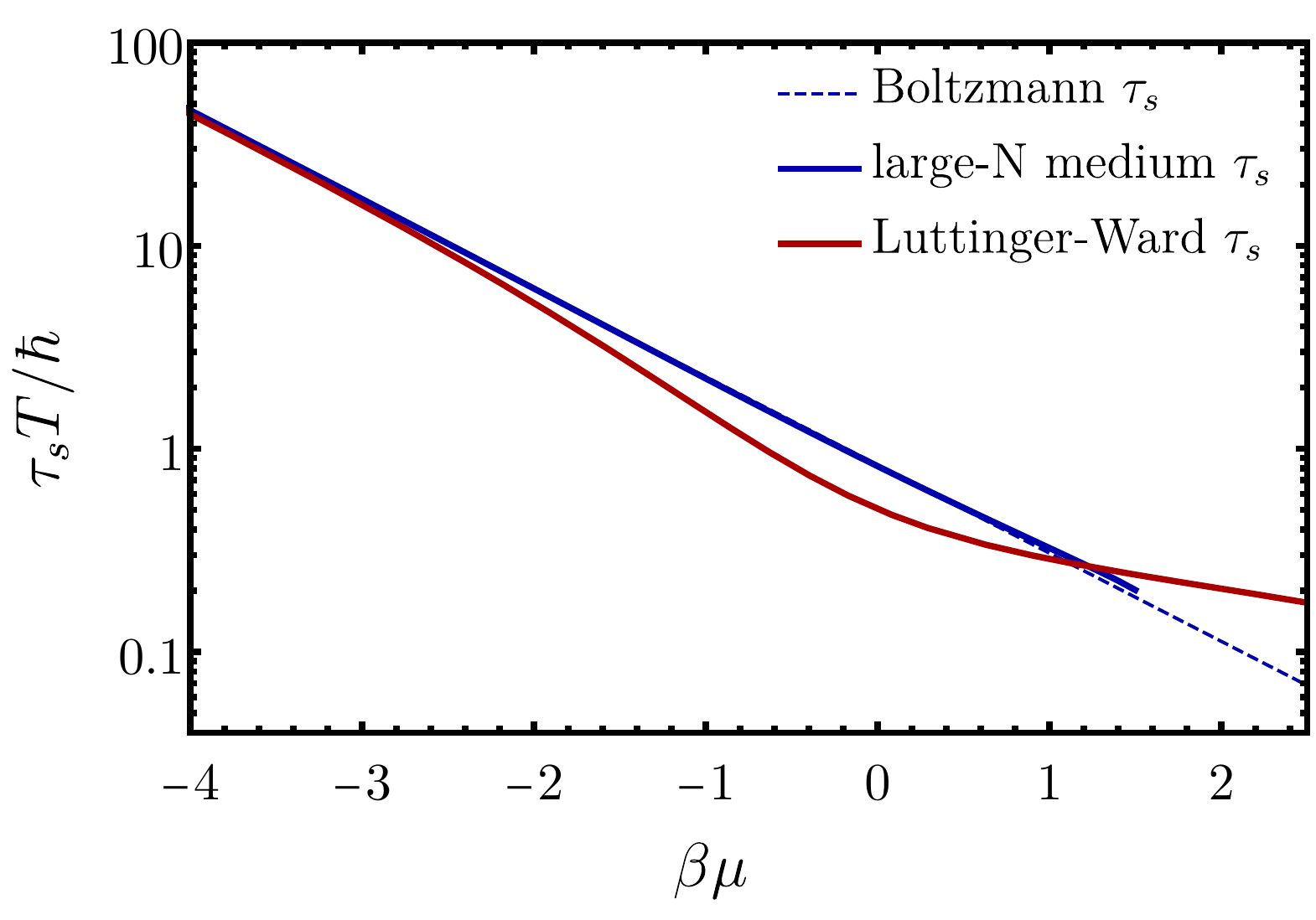}
  \caption{Spin scattering times $\tau_s$ from Boltzmann (dashed) and
    large-$N$ calculations (blue) coincide and agree well with
    Luttinger-Ward results (red) in the quantum critical regime.}
  \label{fig:spintau}
\end{figure}
Finally, also for spin diffusion we compute the transport scattering
time with more than one basis function in the quantum degenerate
regime.  The spin diffusivity $D_s$ is defined via the Einstein
relation in terms of spin conductivity $\sigma_s$ and spin
susceptibility $\chi_s$ \cite{enss2012spin},
\begin{align}
  D_s = \frac{\sigma_s}{\chi_s}
  = \frac{n\tau_s}{m\chi_s}\,.
\end{align}
The spin scattering time $\tau_s$ shown in Fig.~\ref{fig:spintau} also
exhibits the quantum critical scaling that we observed already for
shear and thermal transport: the medium scattering time is, within our
numerical resolution, identical to the Boltzmann scattering time
$\tau_sT/\hbar = \frac{3\pi}{8\sqrt2}e^{-\beta\mu}$.  The quantum
critical scattering time is now combined with the Luttinger-Ward
equation of state for density $n$ and spin susceptibility $\chi_s$ to
obtain the spin diffusivity $D_s$.

\begin{figure}[t]
  \centering
  \includegraphics[width=\linewidth]{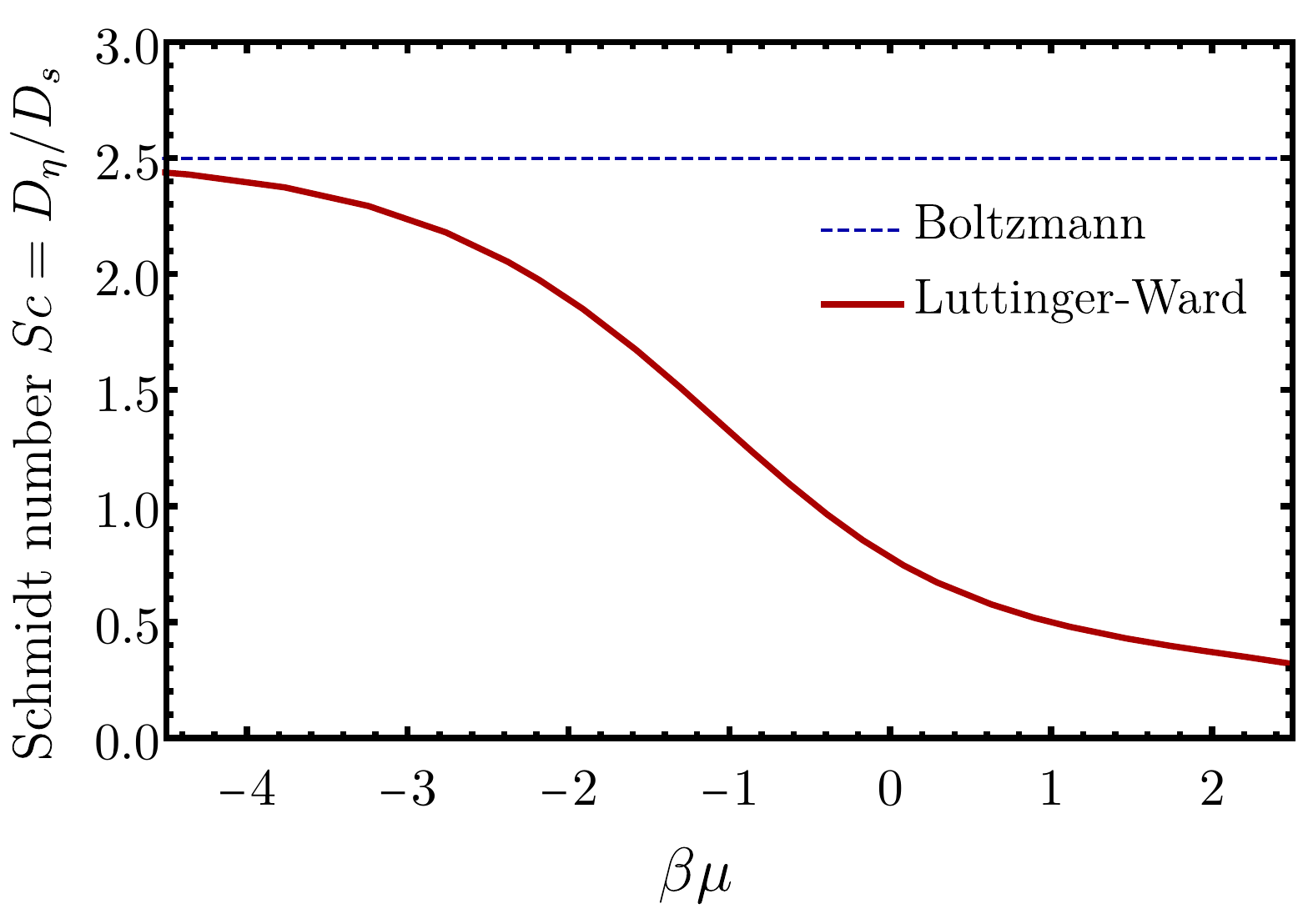}  
  \caption{Schmidt number $\text{Sc}=D_\eta/D_s$ in the quantum
    critical regime, combining quantum critical scattering times from
    our large-$N$ calculation with the Luttinger-Ward equation of
    state.  The dashed line denotes the Boltzmann limit
    $\text{Sc}=5/2$.}
  \label{fig:schmidt}
\end{figure}
Now, the Schmidt number \cite{smith1989}
\begin{align}
  \text{Sc}
  = \frac{D_\eta}{D_s}
  = \frac{(p/mn)\tau_\eta}{(n/m\chi_s)\tau_s}
  = \frac{p\chi_s}{n^2} \times \frac{\tau_\eta}{\tau_s}
\end{align}
is defined as the dimensionless transport ratio of shear and spin
diffusion and characterizes the relative importance of momentum and
spin relaxation.  As shown in Fig.~\ref{fig:schmidt}, the Schmidt
number starts from a value of
$\text{Sc} = \tau_\eta/\tau_s = 5/2$ in the
high-temperature limit and drops to around $0.3$ near $T_c$,
indicating that momentum diffusion is suppressed by a factor of almost
$10$ relative to spin diffusion.  This is physically expected because
viscosity is carried both by single fermions and pairs and therefore
strongly affected by pair fluctuations near $T_c$, whereas pairs carry
no spin current.

\bibliography{all}

\end{document}